\documentclass[12pt]{article}
\usepackage{amssymb}
\usepackage{amsmath}
\usepackage{hyperref}
\usepackage{graphicx}
\usepackage[font=small,labelfont=bf]{caption}
\usepackage{bm}
\usepackage{latexsym}
\usepackage[utf8]{inputenc}
\usepackage{subfig}
\usepackage{caption}
\begin{document}
	\title{\bf The Phase Transition of $4D$ Yang-Mills Charged GB AdS Black Hole with Cloud of Strings}
	\author{ 
		Faramarz Rahmani\thanks{Corresponding author: Email:  faramarz.rahmani@abru.ac.ir} and Mehdi Sadeghi\thanks{Email: mehdi.sadeghi@abru.ac.ir }\hspace{2mm}\\
		{\small {\em Department of Physics, Faculty of Basic Sciences,}}\\
		{\small {\em Ayatollah Boroujerdi University, Boroujerd, Iran}}
	}
	\date{\today}
	\maketitle
	\abstract{In this paper, we present an exact spherically symmetric and Yang-Mills charged AdS black hole solution in the context of $4D$ Einstein-Gauss-Bonnet (EGB) gravity in the presence of a cloud of strings. The regularity of the solution is checked. Thermodynamics of this solution is studied. The critical behavior, the types of phase transitions in canonical ensemble, the Joule-Thomson expansion, the Clapeyron equation and the critical exponents shall be investigated.}
	
	
	
	\section{Introduction} \label{intro}
	
	\indent The theory of general relativity, which was introduced by Albert Einstein in 1916, has achieved significant successes that include the prediction of gravitational waves,  Mercury orbit precession and the discovery of black holes, etc. However, the quantum aspect of gravity and the physics of dark matter and dark energy have not yet been resolved by general relativity. Physicists believe that Einstein's theory of general relativity needs to be modified to address these unresolved issues. Scalar-tensor theories of gravity \cite{Barrabes:1997kk}, \cite{Cai:1996pj}, \cite{Moffat:2005si}, Lovelock theory of gravity \cite{Lovelock:1971yv}, \cite{Lovelock:1972vz}, braneworld cosmology \cite{Cline:2000xn}, \cite{Nihei:2004xv}, and $4D$ Einstein-Gauss-Bonnet (EGB) gravity \cite{Glavan:2019inb}  are examples of modified theories of gravity.
In five dimensional spacetime, EGB gravity does not contribute to gravitational dynamics, as the GB Lagrangian is a total derivative. The contribution of EGB in $4D$ is achieved by rescaling the GB coupling constant, $\alpha \to \frac{\alpha}{D-4}$ and taking the limit $D \to 4$ \cite{Glavan:2019inb}. This idea is similar to the dimensional regularization in quantum field theory. The resulting theory from this concept is known as 4D Einstein-Gauss-Bonnet ($4D$EGB) gravity and has many interesting properties\cite{Yang:2020jno}.\par 
In the framework of string theory, the idea of a cloud of strings arises from the fundamental concept that all particles and interactions in the universe can be described in terms of vibrating strings. These strings can manifest in various vibrational modes, giving rise to the diverse spectrum of particles and forces observed in nature.\par 
The notion of a cloud of strings \cite{Letelier:1979ej, Ghosh:2014pga,Herscovich:2010vr,Lee:2014dha,Mazharimousavi:2015sfo,Graca:2016cbd} suggests a distribution or ensemble of these fundamental strings, potentially interacting and influencing the spacetime geometry and physical phenomena in their vicinity. This concept has been explored in the context of string theory's implications for cosmology, quantum gravity, and the fundamental structure of spacetime.
The dynamics and properties of a cloud of strings in theoretical physics are typically described using mathematical formalisms such as the Nambu-Goto action, which captures the behavior of the strings in spacetime, and the interactions between the strings and the surrounding geometry.
The study of the cloud of strings \cite{Sadeghi:2019muh} is an active area of research in theoretical physics, with implications for understanding the early universe, black hole physics, and the unification of fundamental forces. While the concept of a cloud of strings is theoretical and speculative, it represents a fascinating avenue for exploring the fundamental nature of reality and the potential connections between string theory and observational phenomena. 
Cloud of strings helps us to realize the connections between black holes and string theory, which is an approach to quantum theory of gravity. \par 
Yang-Mills AdS black holes are very important in the study of gravity and the holographic principle. They are solutions of Einstein's equations coupled to a non-Abelian gauge field. 
The thermodynamic properties of these black holes have been widely studied.\par
One of the interesting features of black holes is their thermodynamic behavior. Therefore, we expect them to undergo a phase transition like a real gas. On the other hand according to AdS/CFT correspondence, there is a dictionary between the fields in a bulk of AdS space and its dual conformal field on the boundary the AdS space. Therefore, some field theory problems can be solved through the study of thermodynamic behavior of the bulk.
In the context of black holes, phase transitions can occur in the thermodynamic properties of the black hole, such as its entropy, temperature, pressure and mass.\par 
One well-known example of a phase transition in black holes is the Hawking-Page phase transition, where a transition occurs between a thermal AdS (Anti-de Sitter) black hole and thermal AdS space. This transition is related to the behavior of the black hole in the presence of a negative cosmological constant and has significant implications for the AdS/CFT (Anti-de Sitter/Conformal Field Theory) correspondence \cite{Maldacena}-\cite{Witten:1998qj}  in theoretical physics.
Another example is the phase transition between small and large black holes in the context of black hole thermodynamics. This transition is associated with changes in black hole's entropy and temperature, and it is analogous to the phase transitions observed in other thermodynamic systems.\par 
This paper delves into the intriguing realm of phase transitions in the 4-dimensional (4D) Einstein-Gauss-Bonnet-Yang-Mills theory in the presence of a cloud of strings. Here, we consider all three effects of cloud of strings, the Yang-Mills charge and the Gauss-Bonnet(GB) gravity to have a more comprehensive thermodynamic investigation. By investigating the phase transition phenomena in this extended gravitational model, we aim to unravel its implications for our understanding of the evolution of the universe, as well as its potential connections to contemporary theoretical frameworks such as string theory and dark energy. This exploration not only contributes to the ongoing discourse on gravitational theories but also paves the way for new avenues of research and inquiry in the other fields. This investigation is a significant step in advancing our understanding of the interplay between gravity, gauge fields, and dark energy, and its potential impact on the broader landscape of theoretical physics.\par 
The concept of phase transitions in black holes has been studied extensively in the context of theoretical physics, particularly in relation to the AdS/CFT correspondence, quantum gravity, and the thermodynamics of black holes. This provides valuable insights into the behavior of black holes and inhances our understanding of fundamental physics and the nature of space-time \cite{kubi,sr}.\par 
In the following, we first consider $4D$ Einstein-Gauss-Bonnet-Yang-Mills theory (EGBYM) in the presence of a cloud of strings in section (\ref{sec2}). Then, we shall investigate the thermodynamics and phase transition of the solution in section (\ref{sec3}).
	\section{$4D$ AdS Einstein-Gauss-Bonnet-Yang-Mills Black hole with a Cloud of Strings}
	\label{sec2}
\indent The action for AdS Einstein-Gauss-Bonnet-Yang-Mills in the presence of a cloud of strings is given by:
	\begin{equation}\label{Action}
		I =\frac{1}{16\pi }\int{d^Dx\sqrt{-\mathfrak{g}}\Bigg[R-2 \Lambda+\frac{g}{D-4}\mathcal{G}+F^{(a)}_{\mu \alpha }F^{(a)\mu \alpha}\Bigg]}+\int_{\Sigma} \mu \sqrt{-\gamma} d\tau d\sigma,
	\end{equation}
where $R$ is the scalar curvature, $\mathfrak{g}$, is the determinant of the spacetime metric, $\Lambda=\frac{-(D-1)(D-2)}{2 l^2}$ is the cosmological constant, $g$ is a positive Gauss-Bonnet coupling constant with dimension  $(\text{length})^2$, $\mathcal{G}=R^2-4R_{\mu \nu}R^{\mu \nu}+R_{\mu \nu \rho \sigma }R^{\mu \nu \rho \sigma}$, and $F^{(a)}_{\mu \nu} = \partial_{\mu} A^{(a)}_{\nu} - \partial_{\nu} A^{(a)}_{\mu} - i[A^{(a)}_{\mu}, A^{(a)}_{\nu}]$ is the Cartan subalgebra of the $SU(2)$ Yang-Mills field strength tensor with a gauge coupling constant of 1. Here, $A_{\nu}$'s represent the Cartan subalgebra of the $SU(2)$ gauge group Yang-Mills potentials. The last term is the Nambu-Goto action where $(\tau,\sigma)=({\lambda}^{0},{\lambda}^{1})$ parametrize the worldsheet, $\mu$ is a positive quantity related to the tension of the string, and $\gamma$ is the determinant of the induced metric \cite{Ghosh:2014pga},\cite{Ranjbari:2019ktp}.
We consider the line element as \cite{Ge}
	\begin{equation}\label{metric1}
		ds^{2} =-f(r)dt^{2} +\frac{dr^{2}}{f(r)} +r^2d\Omega^2_{2,\kappa}, \qquad d\Omega^2_{2,\kappa}=\frac{dx^2}{1-\kappa x^2}+x^2 d\varphi^2
	\end{equation}
	
where $\kappa =1 $ for black holes. The blackening function $f(r)$, is defined by a new function $\psi(r)$ as bellow:
	\begin{equation}\label{ff}
		f(r)=1-r^2\psi(r).
	\end{equation}
The equation of motion for the Yang-Mills field is obtained through the variation of action (\ref{Action}) with respect to the field $A^{(a)}_{\mu}$, resulting in:
\begin{equation}\label{EOM-YM}
\nabla_{\mu}F^{(a)\mu \nu} = 0.
\end{equation}
For the solution of Eq.(\ref{EOM-YM}), we adopt the following ansatz for the gauge field \cite{Shepherd:2015dse}:
\begin{equation}\label{background}
	{\bf{A}}^{(a)} =\frac{i}{2}h(r)dt\begin{pmatrix}1 & 0 \\ 0 & -1\end{pmatrix},
\end{equation}
where
\begin{equation}
	h(r)=  C_2+Q\int^{r}\frac{ 1}{u^{D-2}}du,
\end{equation}
and
\begin{equation}
	h'(r)= \frac{ Q}{r^{D-2}}.
\end{equation}
By substituting  ansatz (\ref{metric1}) into the action Eq.(\ref{Action}), we obtain,
	\begin{align}\label{EOMI}
	&I =\frac{\Omega_{D-2}(D-2)}{16\pi }\nonumber\\&\int{dt dr \Bigg[r^{D-1}\psi\bigg(1+g(D-3)\psi\bigg)+\frac{r^{D-1}}{l^2}+\frac{2Q^2 r^{3-D}}{(D-3)(D-2)}-\frac{2 a r}{(D-2)} \Bigg]'},
\end{align}
where $'$ denotes derivative with respect to the $r$. Here, the parameter $a$ is related to the density of cloud of strings and comes from the energy-momentum tensor of the cloud of strings which we have not written the relevant relation here. One can see such a relation in Ref. \cite{Letelier:1979ej}.
 The function $\psi$ is determined by solving for the roots of a quadratic polynomial:
\begin{equation}\label{psi}
	\psi+g(D-3)\psi^2=\frac{16 \psi M}{(D-2)r^{D-1}\Omega_{D-2}}-\frac{1}{l^2}-\frac{2Q^2 r^{4-2D}}{(D-3)(D-2)}-\frac{2a}{(D-2)r^{D-2}},
\end{equation}
where $M$ is a constant.
 By solving Eq.(\ref{psi}) and substituting $\psi$ into Eq.(\ref{ff}), we obtain
\begin{equation}\label{f}
	\begin{split}
		& f(r)=1-\frac{r^2}{2g(D-3)}\Bigg[-1\pm \\ &\sqrt{1-4(D-3)g\bigg(\frac{1}{l^2}+\frac{2Q^2 r^{4-2D}}{(D-3)(D-2)}-\frac{16 \pi M r^{1-D}}{(D-2)\Omega_{D-2}}+\frac{2 a r^{2-D} }{D-2}\bigg)}\Bigg].
	\end{split}
\end{equation}
where $M$ and $Q$ are integration constants proportional to the mass and charge of the black hole, given by the following relations:
\begin{equation}\label{mq}
		M=\frac{(D-2)\Omega_{D-2}}{16 \pi}m
		\end{equation}
		and
		\begin{equation}
		Q^2=\frac{(D-2)(D-3)}{2}q^2.
		\end{equation}
Where, $\Omega_{D-2}=\frac{2 \pi^{\frac{D-1}{2}}}{\Gamma(\frac{D-1}{2})}$ is the surface area of the unit $D-2$ dimensional sphere.
\begin{figure}[h!]
\centering
\subfloat[a]{\includegraphics[width=5cm]{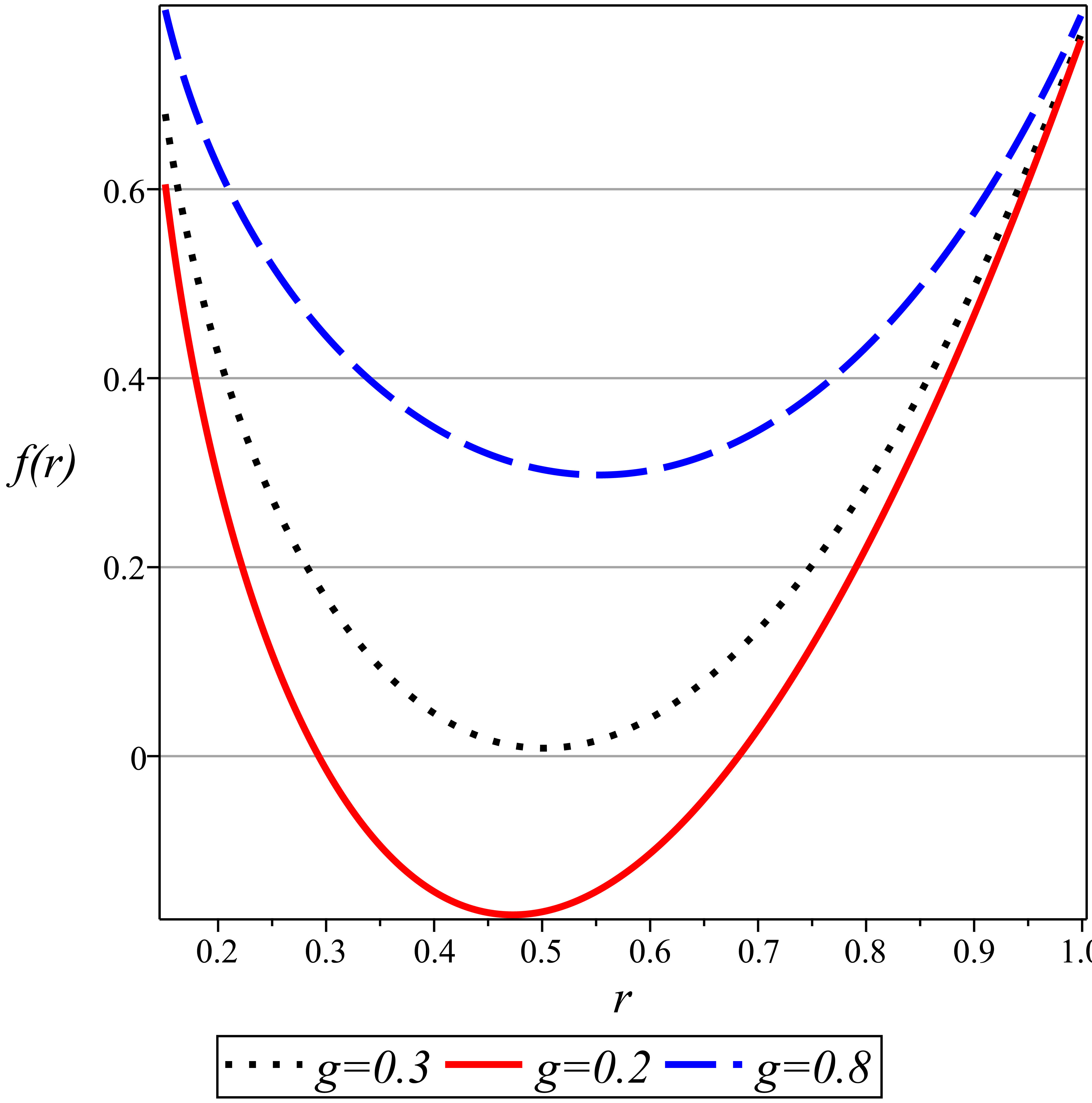}}\quad
\subfloat[b]{\includegraphics[width=5.3cm]{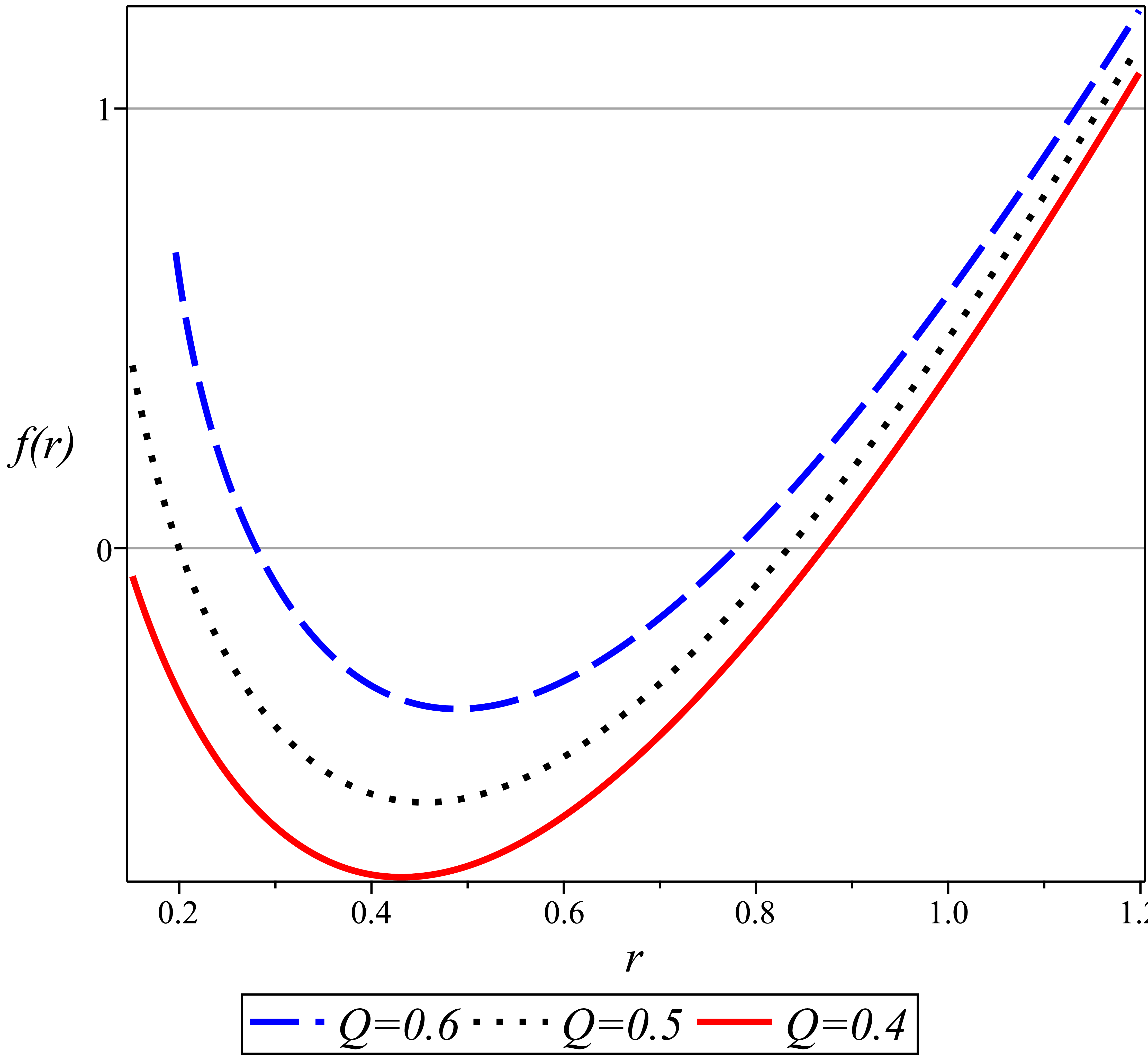}}\quad
\subfloat[b]{\includegraphics[width=5cm]{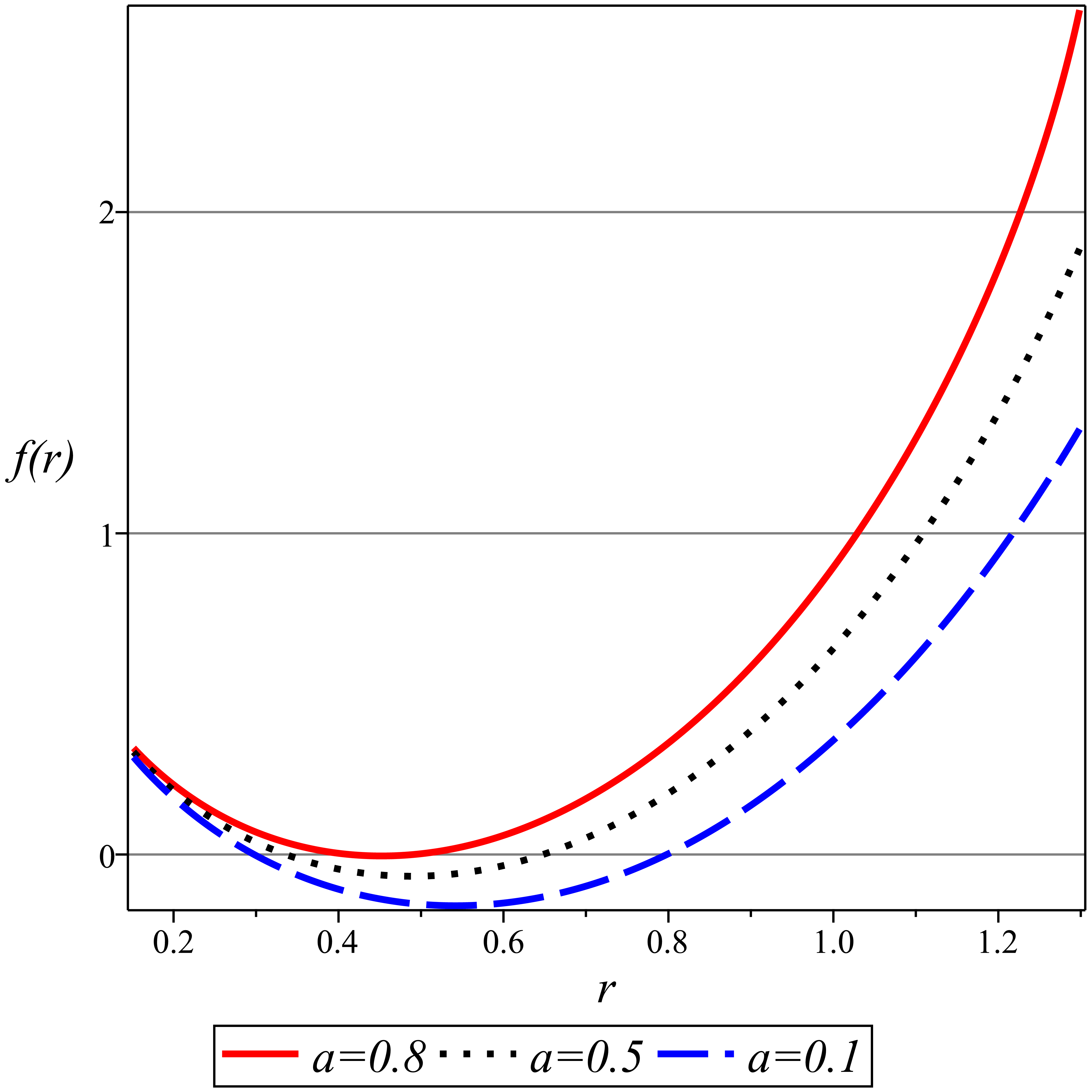}}
\caption{The diagrams of the $f(r)$ for the fixed values $\Lambda=-3$ and $m=1$; \textbf{(a)}: $a=0.5, Q=0.8$, \textbf{(b)}: $a=0.2, g=0.1$ and \textbf{(c)}: $Q=0.3, g=0.4$ \label{fig:0}}
\end{figure}	
In four dimensional spacetime, $M=m$ i.e., the ADM mass of the black hole is equal to the integration constant $m$. 
By defining new parameters as
\begin{equation}
			\frac{(D-3)g}{l^2}=\lambda_{gb}
		\end{equation}
		and
		\begin{equation}
		\frac{2 a  }{D-2}=A,
		\end{equation}
		
Now, by substituting these relations into Eq.(\ref{f}), we get
\begin{equation}\label{f1}
	f(r)=1+\frac{r^2}{2\lambda_{gb} l^2}\Bigg[1\mp \sqrt{1-4\lambda_{gb} \bigg(1+\frac{q^2l^2}{r^{2D-4}}-\frac{ m l^2}{r^{D-1}}+\frac{A l^2}{r^{D-2}}\bigg)}\Bigg].
\end{equation}
The integration constant $m$ is determined by using the condition $f(r_+)=0$. The results in four dimensions are
\begin{equation}\label{ff}
f(r)= 1+\frac {r^{2}}{2g} \left(1-\sqrt{1-4 g \left(-\frac{\Lambda}{3}+\frac{Q^{2}}{r^{4}}-\frac{2 m}{r^{3}}+\frac{a}{r^{2}}\right)}\right)
\end{equation}
and
\begin{equation}
	m=-\frac{r_+^{3} \Lambda}{6}+\frac{ a r_+}{2}+\frac{Q^{2}}{2 r_+}+\frac{r_+}{2}+\frac{g}{2 r_+},
\end{equation}
where, we have chosen the minus sign in the bracket of the relation (\ref{f1}) to obtain the true mass and horizon $r_+$ for subsequent investigations. 
To have a better view of the geometry of the black hole and its horizons, we plot $f(r)$ for different values of the system parameters. See Fig. (\ref{fig:0}). The investigations show that this system  has a maximum of two horizons. We should choose the outer horizon because determines the causal structure of the black hole spacetime and is suitable for studying the thermodynamics of the system.\par
Before studying the thermodynamic behavior of this model, it is valuable to take a look at the singularity or regularity of the solution. For this purpose, we use the Kretschmann scalar, which is defined as follows,
\begin{equation}\label{KS1}
K=R_{\mu \nu \rho \sigma} R^{\mu \nu \rho \sigma}.
\end{equation}
In terms of the metric function $f(r)$ and its derivatives, the Kretschmann scalar takes the form
\begin{equation}\label{KS2}
K=\frac{4}{r^4} \Big( f(r)-1\Big)^2 +\frac{4 f^{\prime}(r)^{2}}{r^2}+f^{\prime \prime}(r)^{2}.
\end{equation}
After substituting relation (\ref{ff}) into the relation (\ref{KS2}), the Kretschmann scalar is given by
\begin{equation}\label{KS3}
\begin{split}
K&=\frac{\left(1-\sqrt{\chi}\right)^{2}}{g^{2}}+\frac{4 \left(\frac{r \left(1-\sqrt{\chi}\right)}{g}+\frac{2r^{2} \mathcal{Y}}{\sqrt{\chi}}\right)^{2}}{r^{2}}\\
&+{\left(\frac{1-\sqrt{\chi}}{g}+\frac{8 r \mathcal{Y}}{\sqrt{\chi}}+\frac{8 r^{2} \mathcal{Y}^{2} g}{\chi^{\frac{3}{2}}}-\frac{2r \mathcal{C}}{\sqrt{\chi}}\right)}^{2}.
\end{split}
\end{equation}
Here, we have defined variables $\chi=1-4g(-\frac{\Lambda}{3}+\frac{Q^2}{r^4}-\frac{2m}{r^3}+\frac{a}{r^2})$, $\mathcal{Y}=-\frac{2Q^2}{r^5}+\frac{3m}{r^4}-\frac{a}{r^3}$ and $\mathcal{C}=-\frac{10 Q^2}{r^5}+\frac{12 m}{r^4}-\frac{3a}{r^3}$. 
It is clear that at $r=0$, the value of the Kretschmann scalar diverges, which is behind the event horizons and is not accessible to us. But, since the variable $\chi$ appears in the denominator of  some terms of relation (\ref{KS3}), we expect that the value of the Kretschmann scalar will diverge at places other than $r=0$. Since we are interested in regular black holes, the parameters involved in the problem should be set in such a way that the divergence points are located at least behind the outer horizon. As an example, consider the blue curve of the middle panel in Fig. (\ref{fig:0}). The values of horizons for this curve with parameters $\Lambda=-3, m=1, Q=0.6,a=0.2$ and $g=0.1$, are $r_{-}=0.280$ and $r_{+}=0.779$ respectively. For these values of parameters, the variable $\chi$ vanishes at $r=0.182$ which is located behind the inner horizon. This can be also examined for the other values of system parameters. 
\section{Thermodynamics and critical behavior}\label{sec3}
In this section, we first investigate the thermodynamic quantities of this model in the extended phase space. Then, $P-V$ critically and phase transition of the system will be studied. As we know, the ADM mass of the AdS black hole plays  the role of enthalpy of the system in the extended phase space. From now, we represent the outer horizon by $r_h$ instead of $r_+$. Thus, according to relation (\ref{mq}) the enthalpy ($H=M=m$) is as bellow
\begin{equation}\label{e1}
\begin{split}
H=&-\frac{r_{h}^{3} \Lambda}{6}+\frac{r_{h} a}{2}+\frac{g}{2 r_{h}}+\frac{r_{h}}{2}+\frac{Q^{2}}{2 r_{h}}=\\
&\frac{4 r_{h}^{3} \pi  P}{3}+\frac{r_{h} a}{2}+\frac{g}{2 r_{h}}+\frac{r_{h}}{2}+\frac{Q^{2}}{2 r_{h}}.
\end{split}
\end{equation}
Where, we have used the relation $\Lambda=-8 \pi P$ in the above equation.
Now, it is appropriate to derive the temperature of the system through the Hawking-temperature relation which relates the black hole temperature to its surface gravity ($\mathcal{K}$) as bellow,
\begin{equation}\label{ht}
\begin{split}
T=\frac{\mathcal{K}(r_h)}{2 \pi}=&\frac{1}{4 \pi} \left(\frac{\partial f(r)}{\partial r}\right)_{r=r_h}=\\
&\frac{-r_{h}^{4} \Lambda +\left(a +1\right) r_{h}^{2}-Q^{2}-g}{4 \pi  r_{h} \left(r_{h}^{2}+2 g \right)}=\\
&\frac{8 r_{h}^{4} \pi  P +\left(a +1\right) r_{h}^{2}-Q^{2}-g}{4 \pi  r_{h} \left(r_{h}^{2}+2 g \right)}.
\end{split}
\end{equation}
Thus, the entropy of the system can be obtained through the relation
\begin{equation}\label{s1}
S=\int_{0}^{r_h} \frac{1}{T} \frac{\partial H}{\partial r_h} dr_h,
\end{equation}
which leads to the
\begin{equation}\label{s2}
S=\pi r_h^2+ 4\pi g \ln(r_h).
\end{equation}
The entropy of the system has been corrected by the GB-term. It is obvious the entropy of the system is not affected by the Yang-Mills charge and a cloud of strings. The correction is due to the modification of geometry. Although, in non-minimal models in which gravitational and matter terms are coupled, the effect of the matter term can be seen in the entropy of the black hole \cite{Sara1}.\par  
Now, the first law of thermodynamics in extended phase space can be written as
\begin{equation}\label{termo}
dH(S,P,Q,a,g)=TdS+VdP+\phi dQ+\mathcal{A}da+\mathcal{B}dg,
\end{equation}
with
\begin{equation}\label{v}
V=\left(\frac{\partial H}{\partial P}\right)_{S,Q,a,g}=\frac{4 \pi}{3}r_h^3,
\end{equation}
\begin{equation}
\phi=\left(\frac{\partial H}{\partial Q}\right)_{S,P,a,g}=\frac{Q}{r_h},
\end{equation}
\begin{equation}
\mathcal{A}=\left(\frac{\partial H}{\partial a}\right)_{S,P,Q,g}=\frac{1}{2}r_h
\end{equation}
and
\begin{equation}\label{GP}
\mathcal{B}=\left(\frac{\partial H}{\partial g}\right)_{S,P,Q,a}=\frac{1}{2 r_h}.
\end{equation}
Here, $V$ is thermodynamic volume conjugate to the pressure  of the black hole. Also, quantities $\phi$, $\mathcal{A}$ and $\mathcal{G}$ are potentials associated to the Yang-Mills charge, cloud of strings and the Gauss-Bonnet terms respectively.
Smarr relation takes the form
\begin{equation}\label{smarr}
H=2TS-2PV+Q\phi+a \mathcal{A}+g \mathcal{B},
\end{equation}
where can be easily checked.
Now, by using relation (\ref{ht}),we write the equation of state as bellow
\begin{equation}\label{state}
P=\frac{T}{2 r_{h}}+\frac{g T}{r_{h}^{3}}-\frac{a+1}{8 \pi  r_{h}^{2}}+\frac{Q^{2}}{8 \pi  r_{h}^{4}}+\frac{g}{8 \pi  r_{h}^{4}}
\end{equation}
First, we have to check whether the system has a critical point or not. For this purpose, we use the condition
\begin{equation}\label{cr1}
	\frac{\partial P}{\partial r_h}|_{T=T_c,r_h=r_c}=0, \qquad  \frac{\partial^2 P}{\partial^2 r_h}|_{T=T_c,r_h=r_c}=0
\end{equation}
which is widely used in this context.
Through the condition (\ref{cr1}), the critical values of the thermodynamic quantities $r_c, P_c$ and $T_c$  are related to the system parameters.
The critical value of the horizon is given by
\begin{equation}\label{rc}
r_c=\sqrt{\frac{3 Q^{2}+3 a g +\mathrm{\Upsilon}+6 g}{a +1}}.
\end{equation}
\begin{figure}[h!]
\centering
\subfloat[a]{\includegraphics[width=5cm]{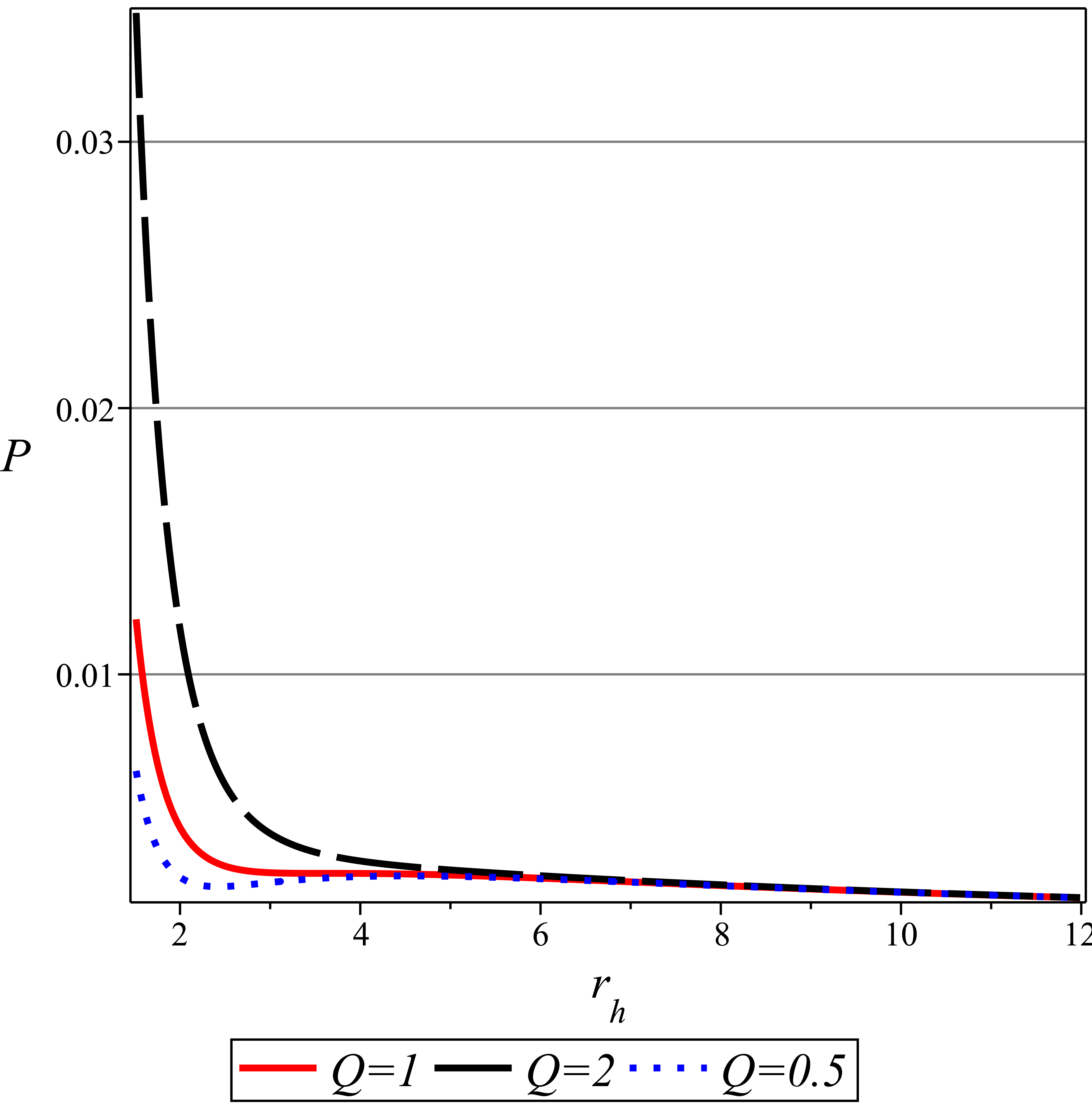}}\quad
\subfloat[b]{\includegraphics[width=5cm]{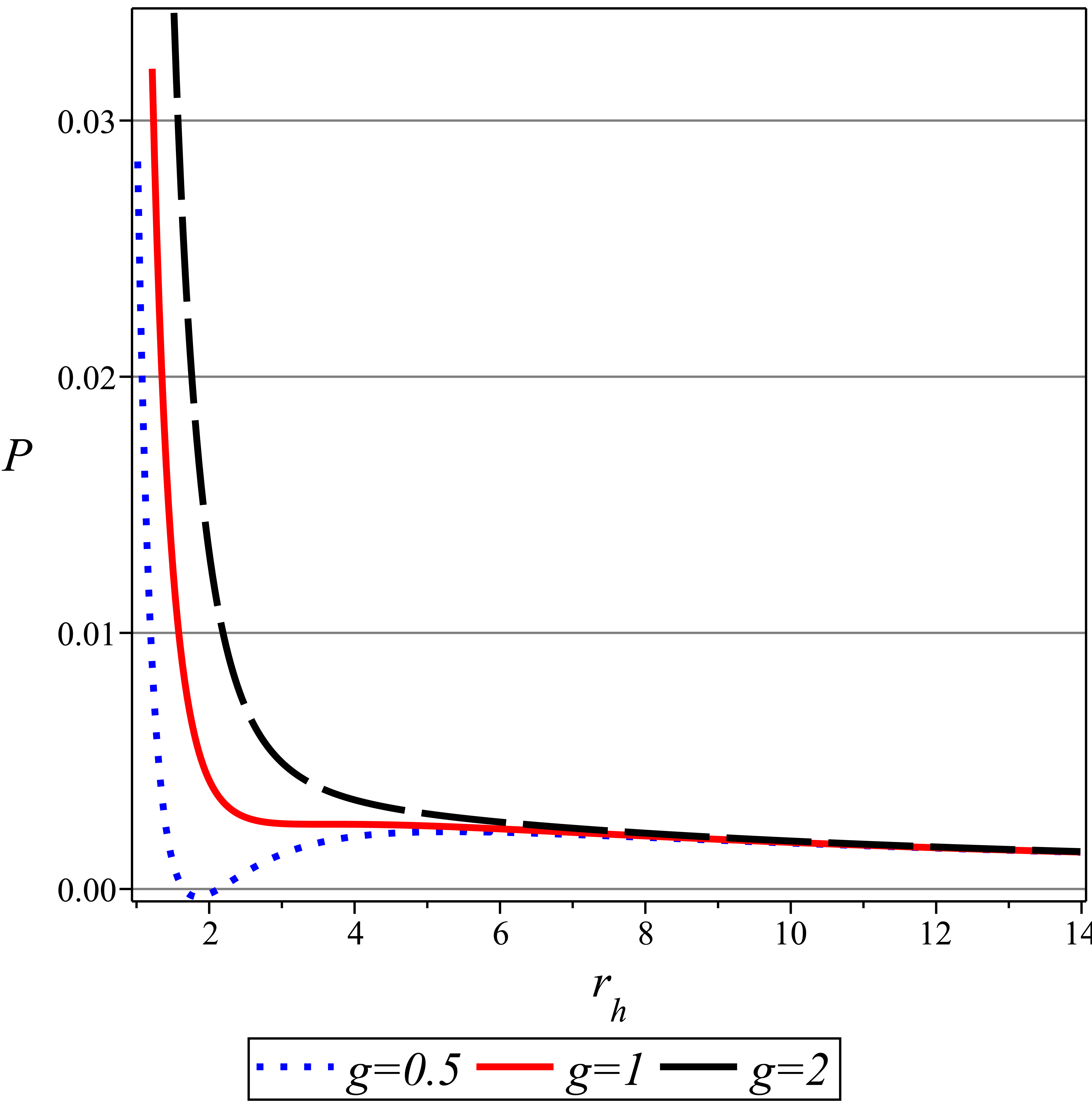}}\quad
\subfloat[b]{\includegraphics[width=5cm]{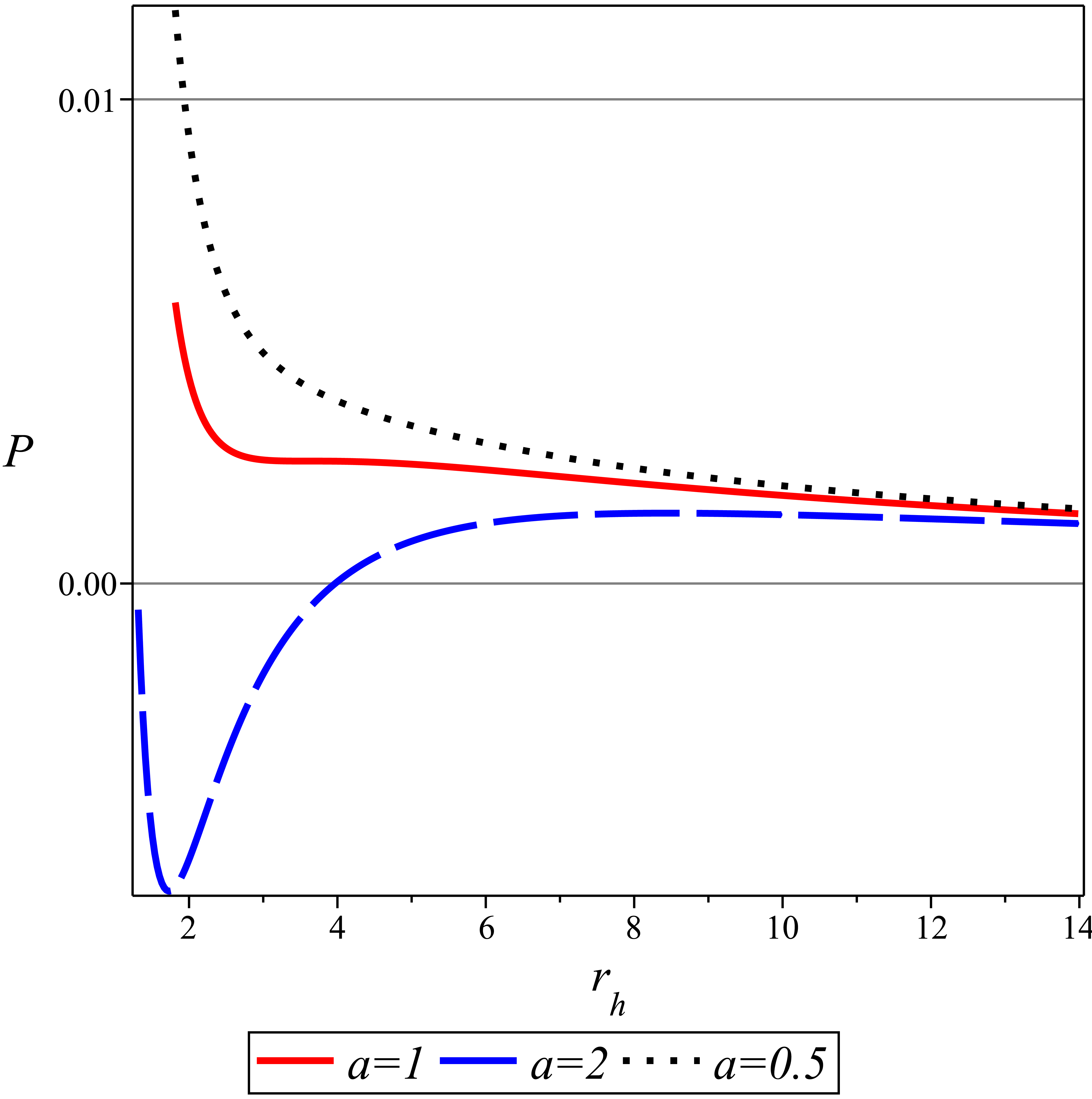}}
\caption{$P-r_{h}$ diagram of the system. The solid lines represent the critical isotherms which are associated to the parameters $a=1, g=1, Q=1$ with $T_c=0.0511$; \textbf{(a)}: the Yang-Mills charge changes, \textbf{(b)}: the GB parameter changes and \textbf{(c)}: the cloud of strings parameter changes. \label{fig:1}}
\end{figure}
The critical pressure and temperature in terms of parameters of the model are in the form
\begin{equation}\label{pc}
P_c=\frac{\left(Q^{2} a g +3 a^{2} g^{2}+3 \mathrm{\Upsilon} Q^{2}+4 \mathrm{\Upsilon} a g +Q^{2} g +7 a \,g^{2}+\mathrm{\Upsilon}^{2}+7 \mathrm{\Upsilon} g +4 g^{2}\right) \left(a +1\right)^{2}}{8 \pi  \left(3 Q^{2}+3 a g +\mathrm{\Upsilon}+6 g \right)^{2} \left(3 Q^{2}+9 a g +\mathrm{\Upsilon}+12 g \right)}
\end{equation}
and
\begin{equation}\label{tc}
T_c=\frac{\left(Q^{2}+3 a g +\mathrm{\Upsilon}+4 g \right) \left(a +1\right)}{2 \pi  \sqrt{\frac{3 Q^{2}+3 a g +\mathrm{\Upsilon}+6 g}{a +1}}\, \left(3 Q^{2}+9 a g +\mathrm{\Upsilon}+12 g \right)}
\end{equation}
respectively. Where, $\Upsilon=\sqrt{9 Q^{4}+30 Q^{2} a g +9 a^{2} g^{2}+48 Q^{2} g +48 a \,g^{2}+48 g^{2}}$.
\begin{figure}[h!]
\centering
\subfloat[a]{\includegraphics[width=5cm]{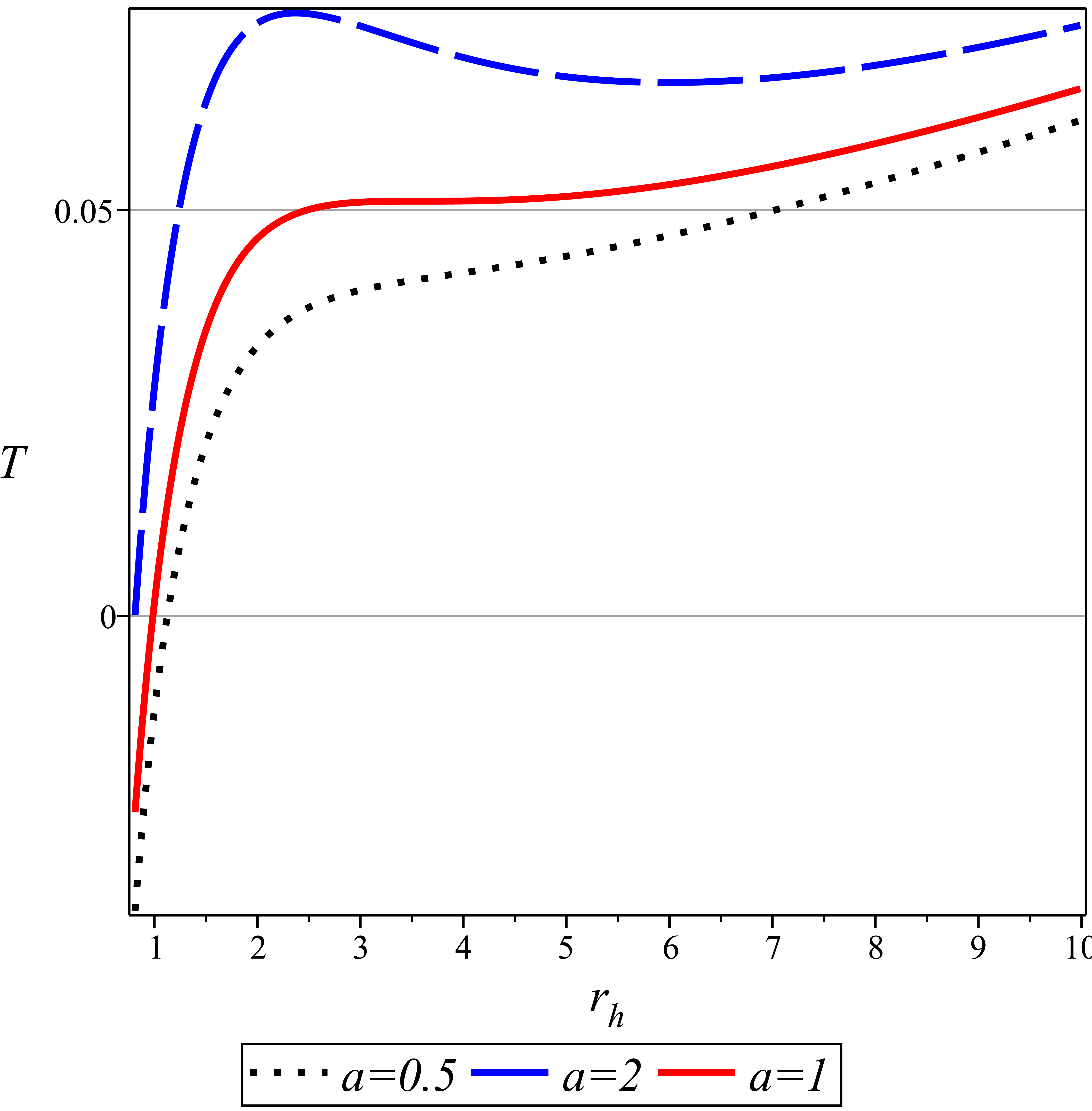}}\quad
\subfloat[b]{\includegraphics[width=5cm]{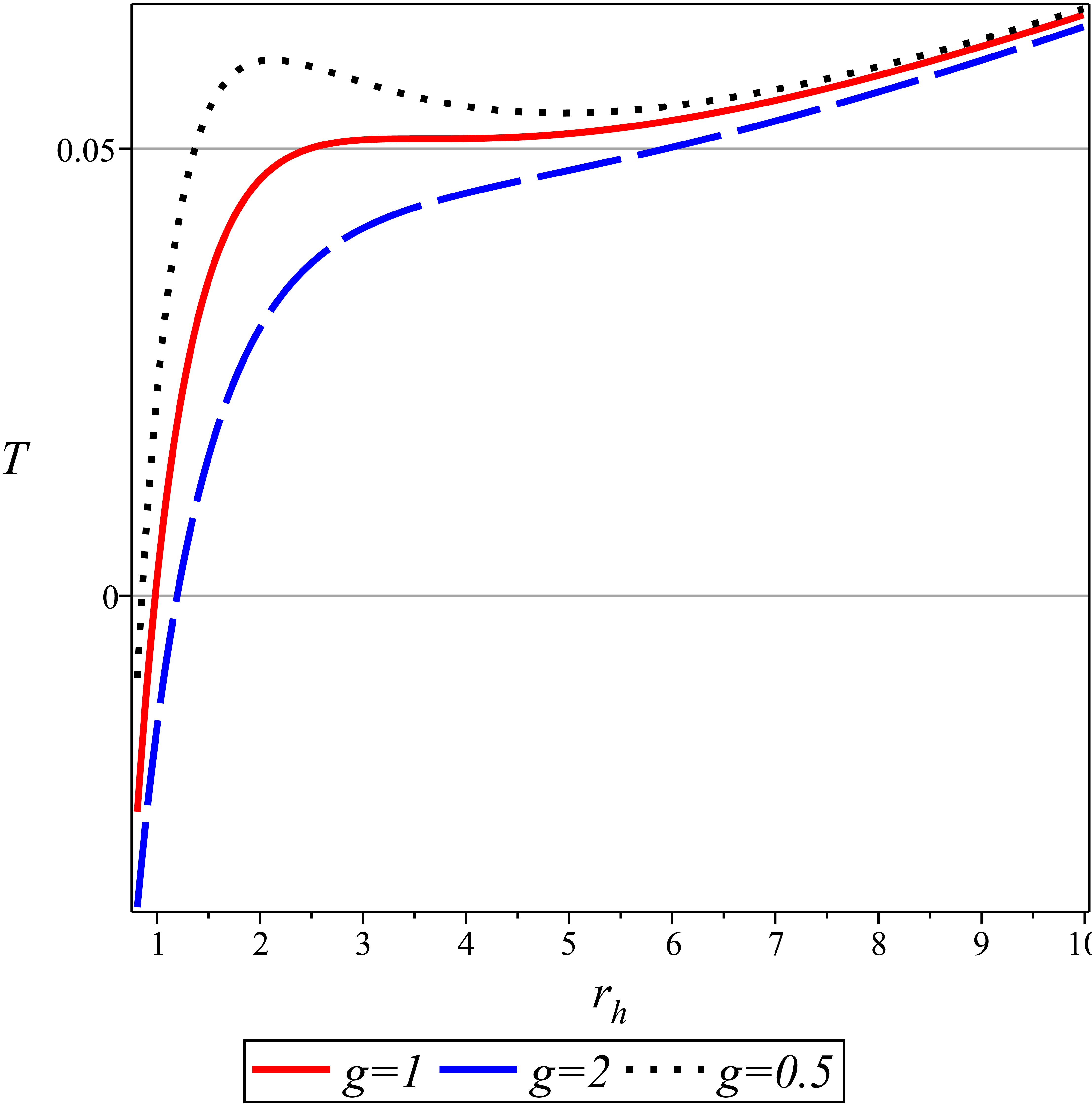}}\quad
\subfloat[b]{\includegraphics[width=5cm]{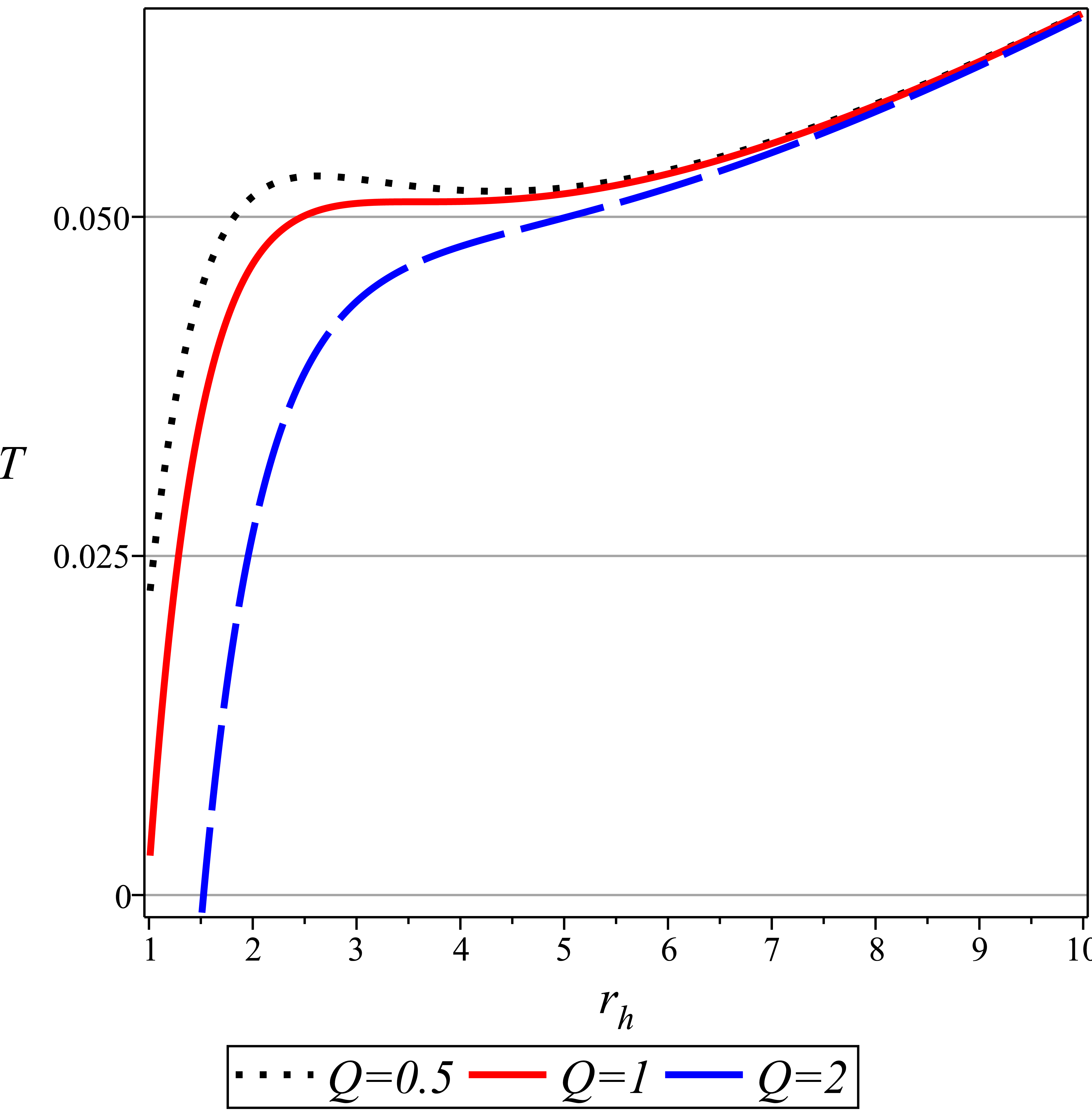}}
\caption{$T-r_{h}$ diagram of the system. The solid lines represent the critical behavior of the system associated to the parameters $a=1, g=1$ and $ Q=1$ with $P_c=0.0025$; \textbf{(a)}: the cloud of strings parameter changes, \textbf{(b)}: the GB parameter changes and \textbf{(c)}: the Yang-Mills charge changes. \label{fig:2}}
\end{figure}
In Fig (\ref{fig:1}) the solid lines show the critical behavior of the system for the values $a=1,g=1$ and $Q=1$. The left panel shows the pressure of the system versus the horizon with change in the value of the Yang-Mills charge. In the middle panel the Gauss-Bonnet parameter, and in the right panel the cloud of strings parameter varies. These, diagrams show the critical behavior in this system. By increasing the Yang-Mills and GB parameters, the behavior of the system tends to the behavior of an ideal gas. While for the cloud of strings parameter, the situation is the opposite.  
The diagrams of the temperature versus horizon also show critical behavior. We can see that at a special radius, $T=0$ which refers to a black hole with zero temperature and nonzero entropy. Solid lines show the critical behavior of the system for different values of parameters. See Fig. (\ref{fig:2}).\par
\begin{figure}[h!]
\centering
\subfloat[a]{\includegraphics[width=5cm]{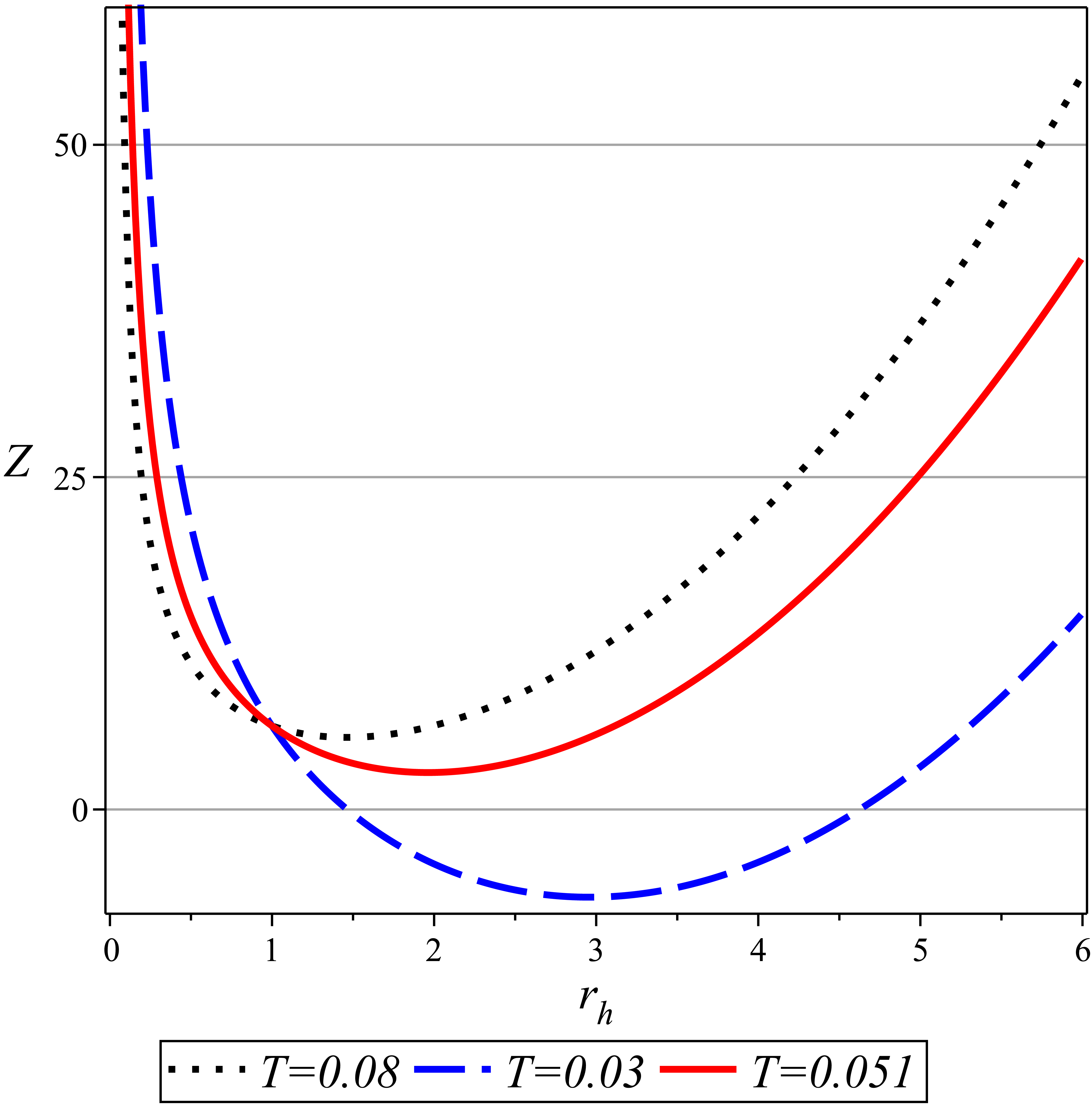}}\quad
\subfloat[b]{\includegraphics[width=5cm]{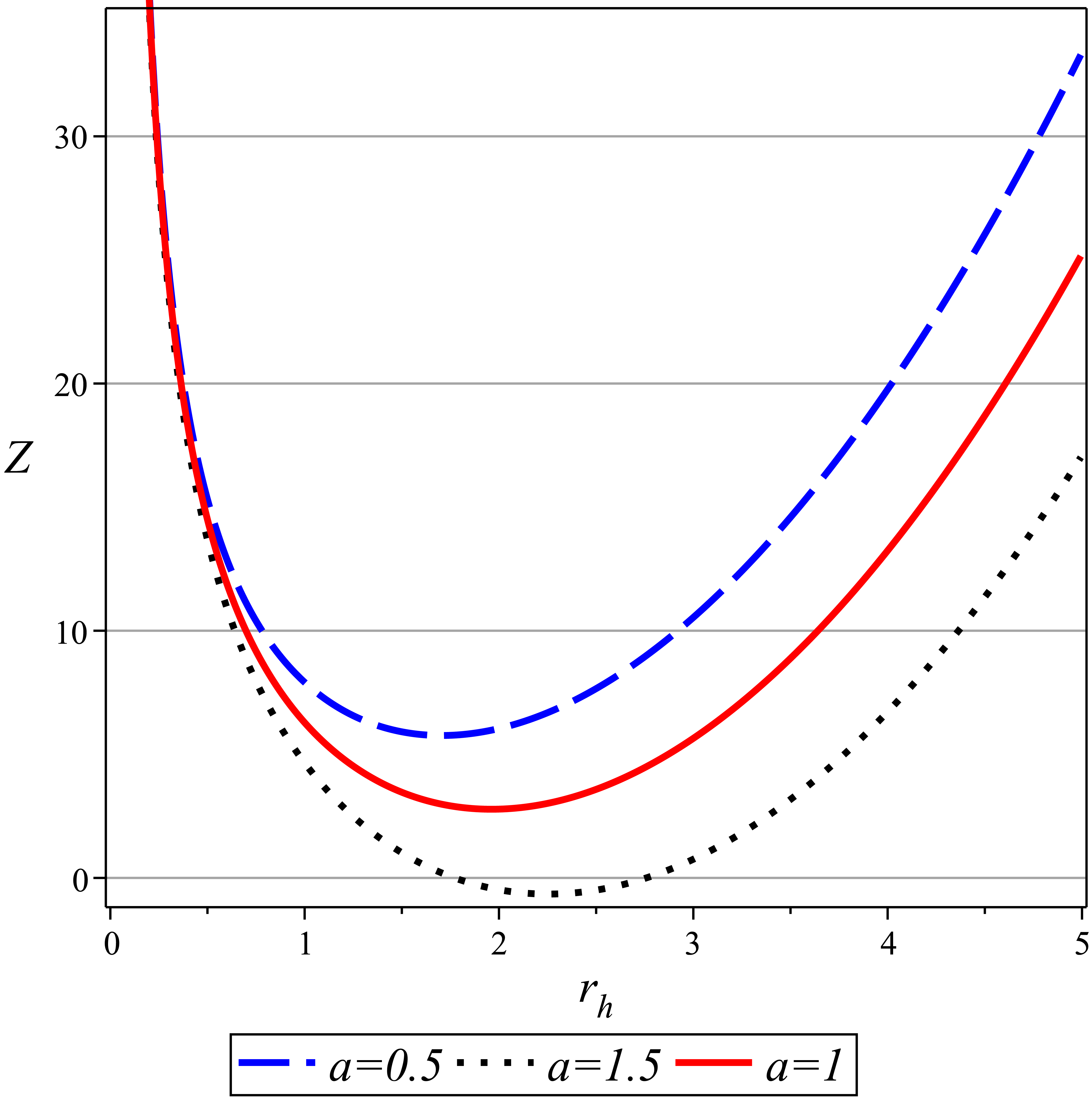}}\quad
\caption{$Z-r_{h}$ diagram of the system ; \textbf{(a)}: the temperature of the system changes and the values of associated  parameters are $a=1, g=1$ and $ Q=1$ respectively,  \textbf{(b)}: the system temperature is fixed at critical temperature $T_c=0.051$, and the cloud of strings parameter changes.  \label{fig:z1}}
\end{figure}

\begin{figure}[h!]
\centering
\subfloat[a]{\includegraphics[width=5cm]{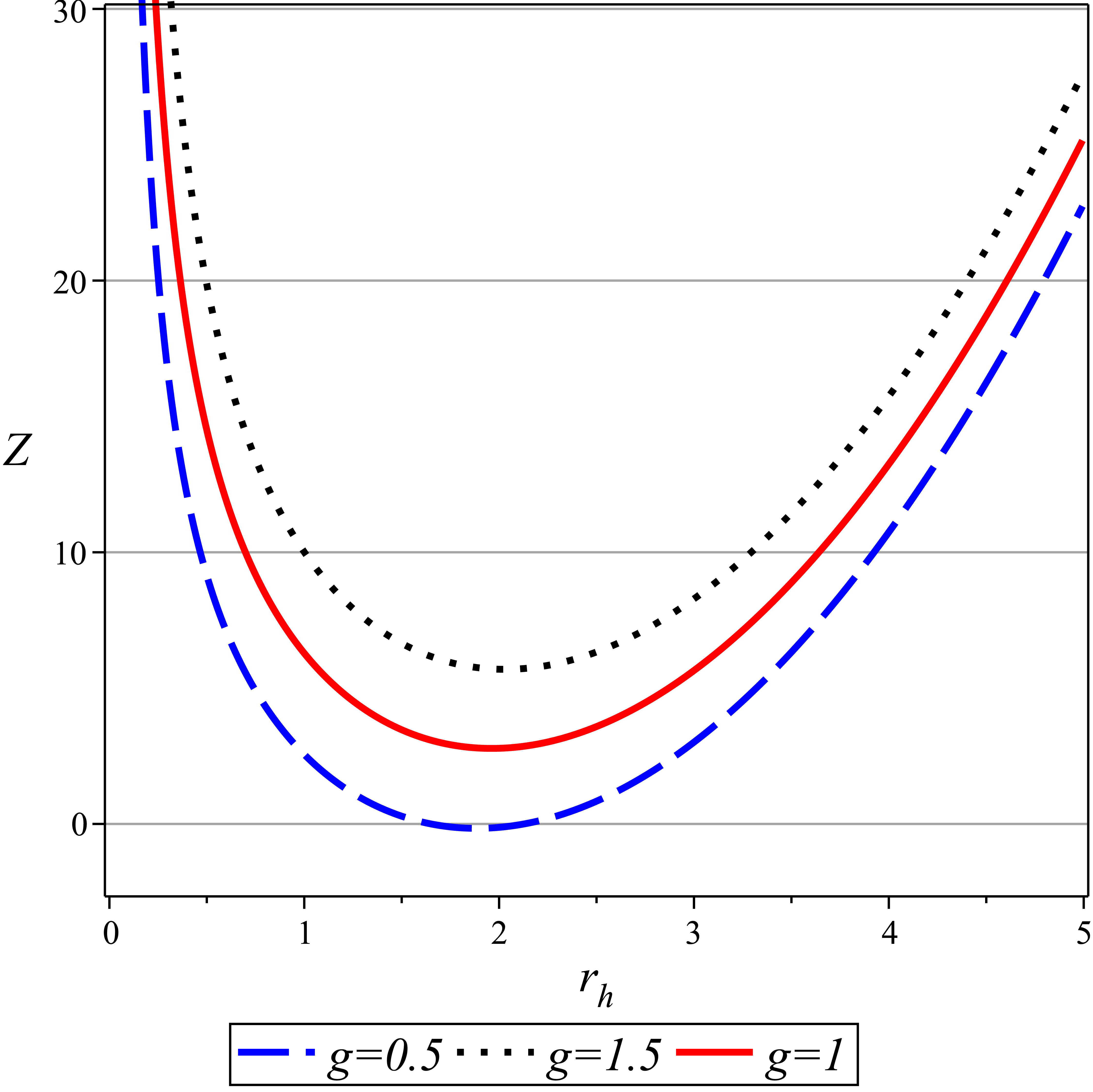}}\quad
\subfloat[b]{\includegraphics[width=5cm]{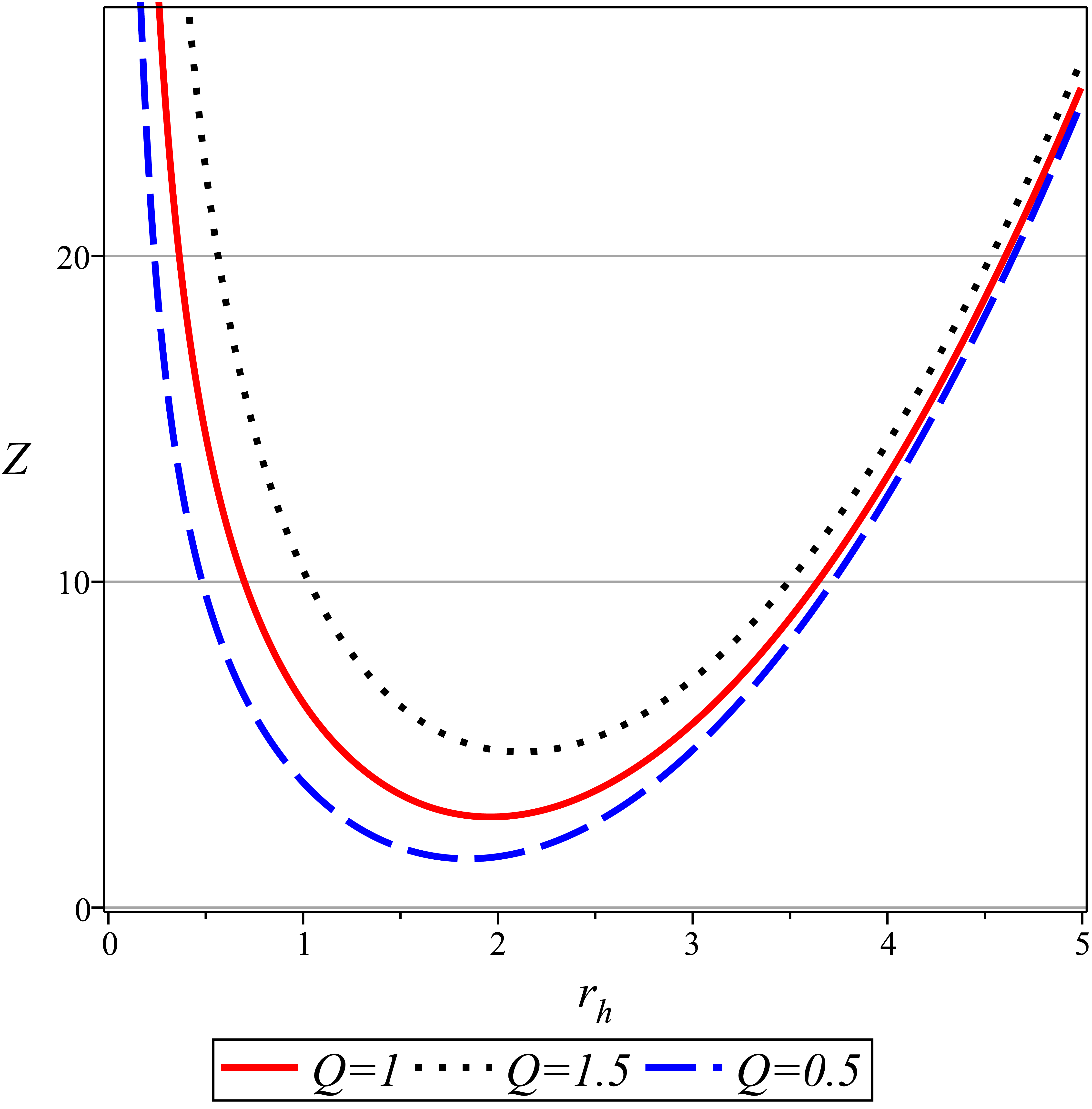}}\quad
\caption{$Z-r_{h}$ diagram of the system for fixed temperature $T_c=0.051$; \textbf{(a)}: the GB parameter changes,  \textbf{(b)}: the Yang-Mills charge changes.  \label{fig:z2}}
\end{figure}

\begin{figure}[h!]
\centering
\subfloat[a]{\includegraphics[width=5cm]{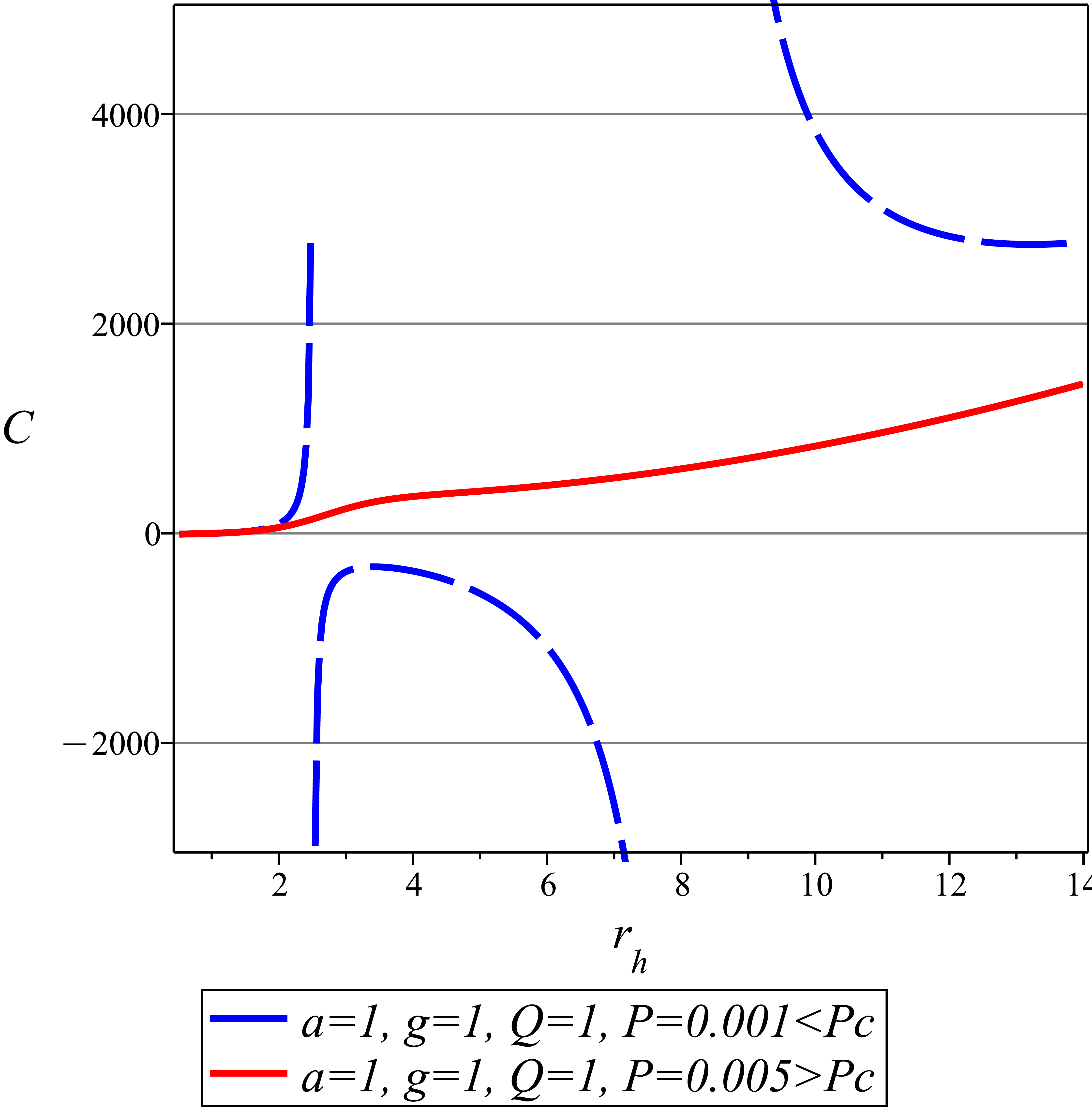}}\quad
\subfloat[b]{\includegraphics[width=5cm]{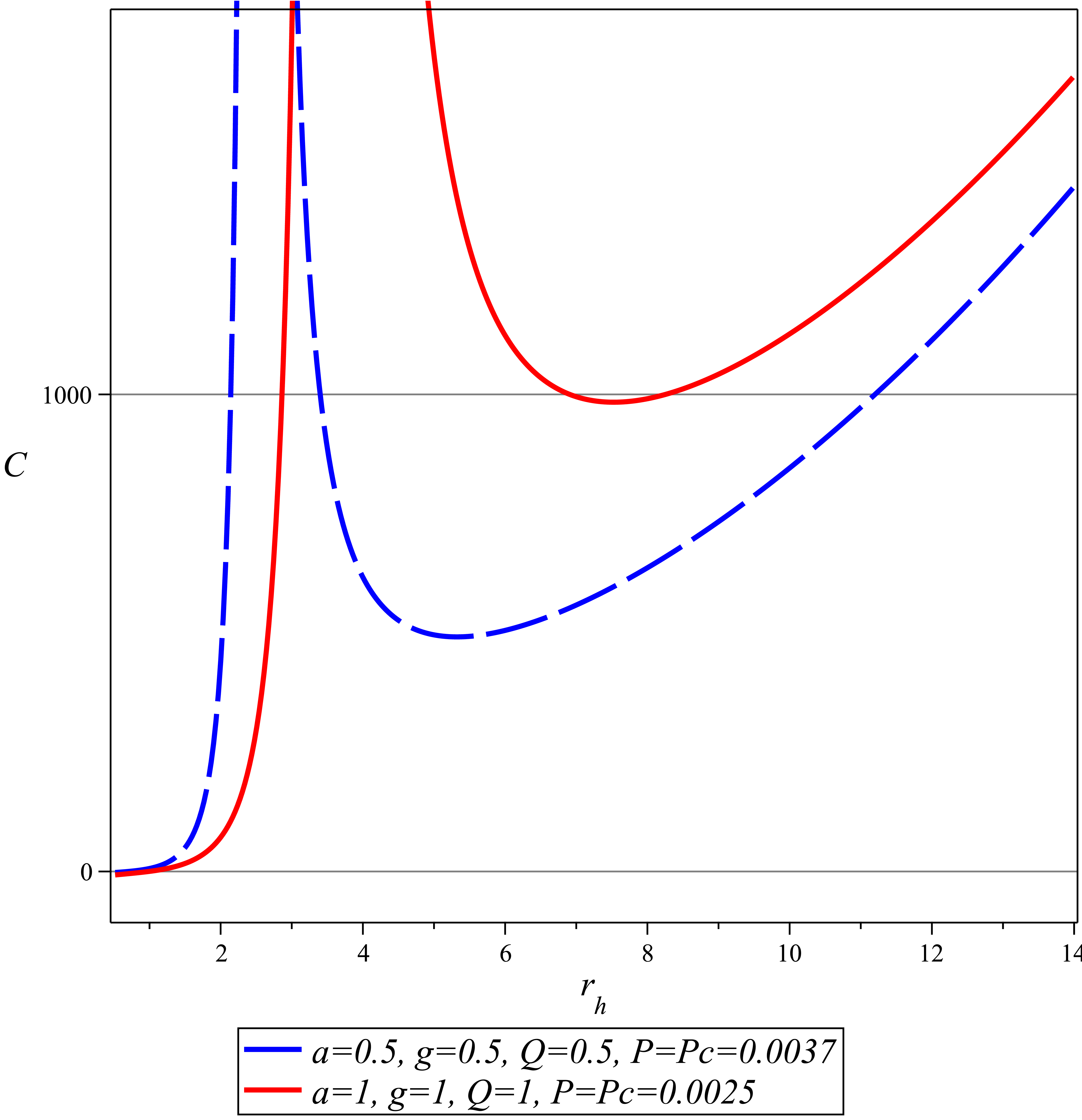}}
\caption{$C-r_{h}$ diagrams of the system; \textbf{(a)}: for pressures higher and less than critical pressure, \textbf{(b)}: for two arbitrary critical pressures of the system. \label{fig:3}}
\end{figure}

 
One of the quantities which measures the deviation of gas behavior from the behavior of an ideal gas is the compressibility factor which is given by
\begin{equation}\label{com}
Z=\frac{PV}{T}.
\end{equation}
For an ideal gas $Z=1$ which shows there is no interaction between particles. For a Van-der Walls gas, the compressibility factor at critical point is independent of the gas parameters and it is equal to $\frac{3}{8}$. While, here $Z$ is a function of system parameters. When, $r_h \longrightarrow \infty$, $Z\longrightarrow \infty$ which shows a different behavior than real gasses. By using relation (\ref{com}), One can show that the compressibility factor has a minimum for different values of the system parameters. The behavior of  compressibility factor versus the black hole horizon can be seen in Figs. (\ref{fig:z1}) and (\ref{fig:z2}).
The values of the compressibility factor at critical points for several arbitrary values of the parameters are seen in the following table. An interesting point is that by increasing the cloud of strings parameter, the compressibility factor at critical point decreases, while by increasing the Yang-Mills charge and GB parameter, increases.
\begin{table}[ht]
\caption{The behavior of the compressibility factor at critical points.} 
\centering 
\begin{tabular}{c c c c c c c} 
\hline\hline 
$a$ & $Q$ & $g$ &$P_c$ & $T_c$ & $r_c$ & $Z_c$ \\ [0.5ex] 
\hline 
0.5 \,\,\,\ & 0.5\,\,\,\ & 0.5\,\,\,\ &0.0037\,\,\,\ &0.0542\,\,\,\ & 2.542& 4.815 \\ 
1\,\,\,\ & 0.5\,\,\,\ & 0.5\,\,\,\ &0.0056\,\,\,\ &0.0758\,\,\,\ &2.376 & 4.174 \\
1.5\,\,\,\ & 0.5\,\,\,\ & 0.5\,\,\,\ &0.0075\,\,\,\ &0.0978 &2.269\,\,\,\ & 3.782 \\
0.5\,\,\,\ & 1\,\,\,\ & 0.5\,\,\,\ &0.0027\,\,\,\ &0.0464 & 3.102\,\,\,\ & 7.296 \\
0.5\,\,\,\ & 1.5\,\,\,\ & 0.5\,\,\,\ &0.0018\,\,\,\ & 0.0388\,\,\,\ &3.840 & 11.319 \\
0.5\,\,\,\ & 0.5\,\,\,\ & 1\,\,\,\ &0.0020\,\,\,\ & 0.0395\,\,\,\ &3.441 & 8.778 \\
0.5\,\,\,\ & 0.5\,\,\,\ & 1.5\,\,\,\ &0.0013\,\,\,\ &0.0326\,\,\,\ &4.149 & 12.738\\  [1ex] 
\hline 
\end{tabular}
\label{table:zz} 
\end{table}
One of the important quantities in thermodynamics which is studied to investigate the phase transition of the is the heat capacity at constant pressure. A divergence in heat capacity can indicate a  second order phase transition near the critical points\cite{Huang,Callen}. Because, the heat capacity is proportional to the second derivative of the Gibbs free energy with respect to the temperature. For stable regions $C_p>0$ and for unstable ones, $C_p<0$. In this model, the heat capacity is given by
\begin{equation}\label{cp}
\begin{split}
C_P&=\left(\frac{\partial H}{\partial r_h}\right)\left(\frac{\partial r_h}{\partial T}\right)=\\
&\frac{2 \left(r_{h}^{2}+2 g \right)^{2} \left(8 r_{h}^{4} \pi  P +r_{h}^{2} a -Q^{2}+r_{h}^{2}-g \right) \pi}{8 \pi  P r_{h}^{6}+48 \pi  g P r_{h}^{4}-a r_{h}^{4}+3 Q^{2} r_{h}^{2}+2 a g r_{h}^{2}-r_{h}^{4}+2 Q^{2} g +5 r_{h}^{2} g +2 g^{2}}
\end{split}
\end{equation}
The divergence points of the heat capacity usually signal the second order phase transition near the critical points of the system. For example, for the values $Q=1, g=1$ and $a=1$, the critical horizon is $r_c=3.5956$ and heat capacity diverges at $r_d=3.5823$. Then, the difference is $\delta=\mid r_c-r_d\mid\simeq 0.01$ which is a small value. The left panel of the Fig. (\ref{fig:3}), shows the heat capacity for two pressures higher and lower than the critical pressure for values $a=1,Q=1$ and $g=1$. It is obvious that by increasing the pressure, the heat capacity tends to be more stable and vice versa. For the pressure $P=0.001<P_c$, the diagram consists of three regions which correspond to small, medium and large black holes respectively. In the right panel the heat capacity has been depicted for two critical pressures associated to parameters $a=1,Q=1,g=1$ and $a=0.5,Q=0.5,g=0.5$. For critical pressures the intermediate regions disappear. 
\begin{figure}[h!]
\centering
\subfloat[a]{\includegraphics[width=5cm]{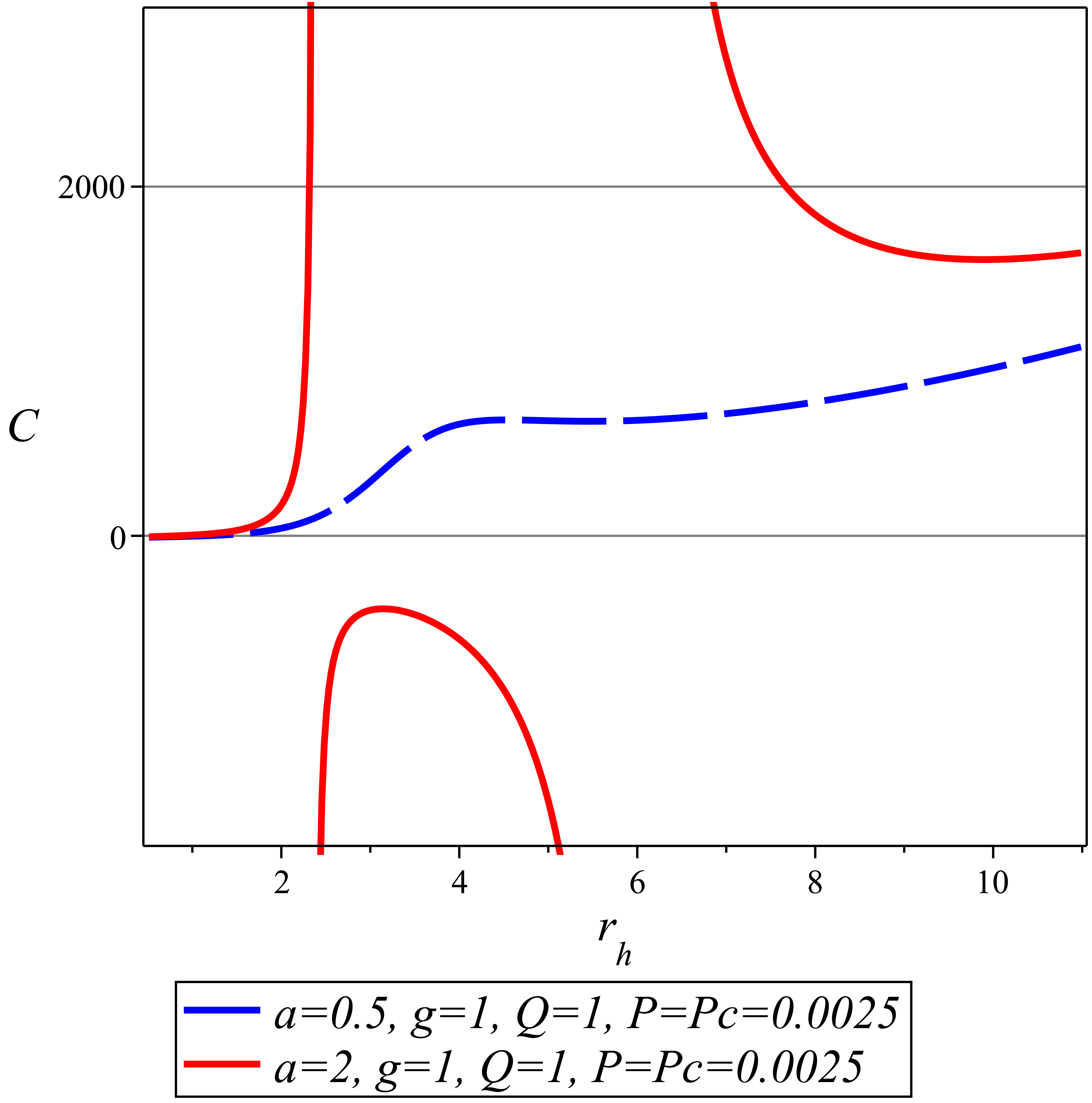}}\quad
\subfloat[b]{\includegraphics[width=5cm]{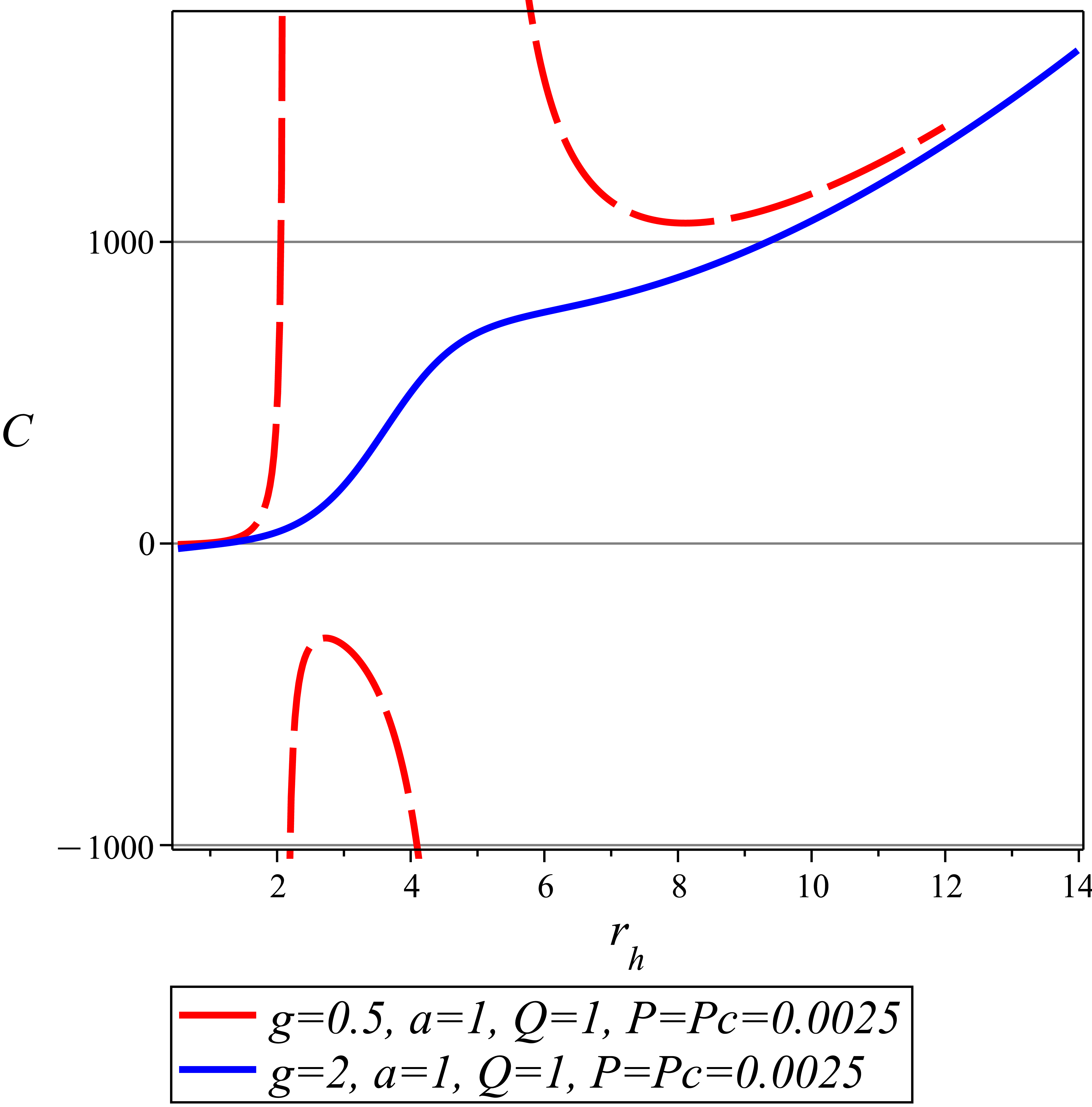}}\quad
\subfloat[b]{\includegraphics[width=5cm]{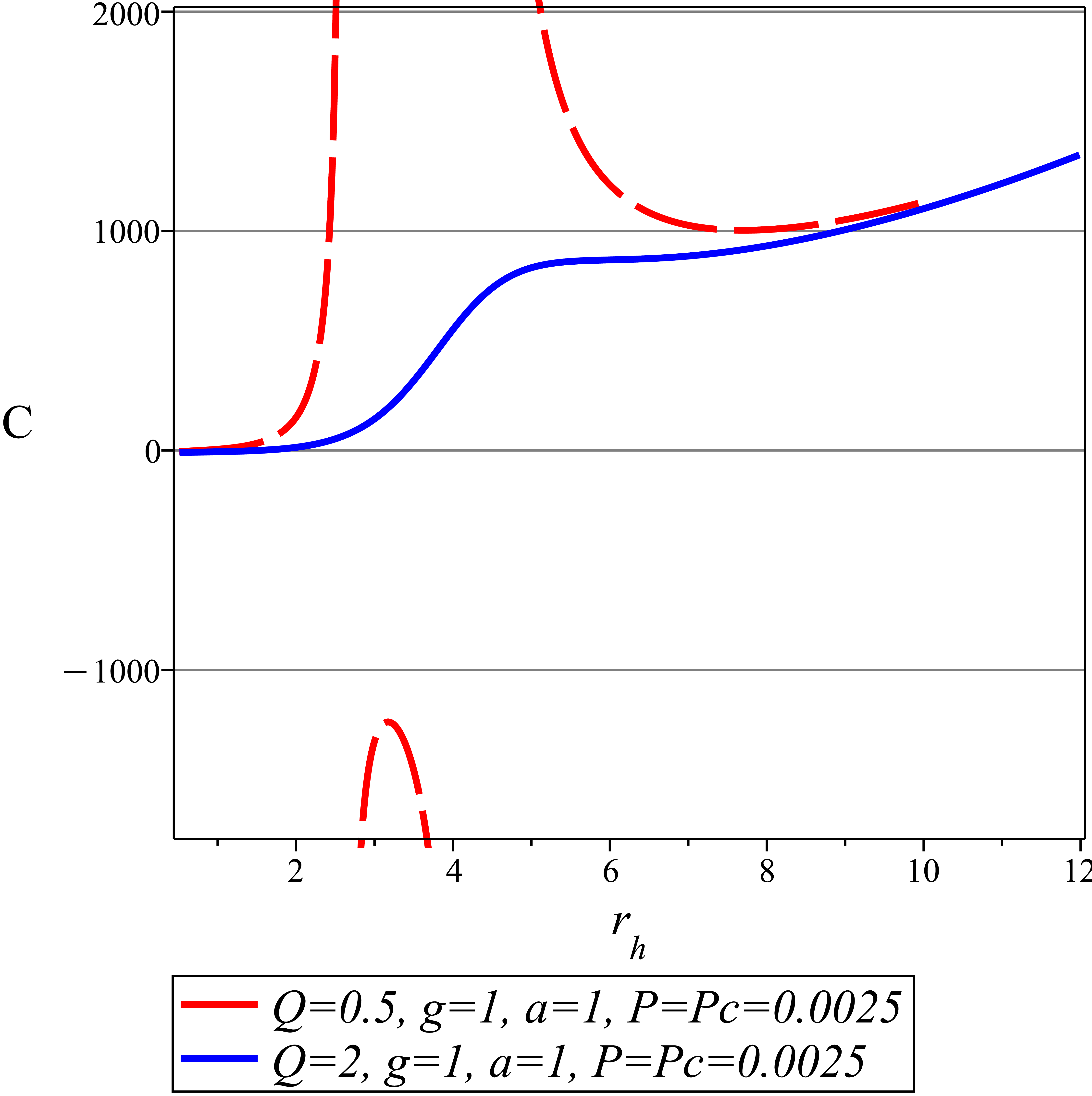}}
\caption{$C-r_{h}$ diagrams of the system; \textbf{(a)}: the cloud of strings parameter changes, \textbf{(b)}: the GB parameter changes and \textbf{(c)}: the Yang-Mills parameter changes. \label{fig:4}}
\end{figure}
In Fig. (\ref{fig:4}), the behavior of the heat capacity for different values of parameters has been illustrated. The left panel shows that for the higher values of the cloud of strings parameter, the heat capacity has an unstable intermediate region and for smaller values the behavior of heat capacity is more stable. While, for the other two parameters, the higher values make the system more stable. The unstable intermediate region with negative heat capacity shows that black hole becomes hotter when it losses energy and vice-versa.\par  
Now, we deal with the Gibbs free energy. 
It is expected that during the first order phase transition  a swallow-tail shape to be seen in the Gibbs free energy diagram of a van-der Walls fluid . For the pressures higher than the critical pressure the Gibbs free energy has smooth behavior. While, for the pressures less than the critical pressure, a swallow-tail behavior is seen. In a first order phase transition a talent heat is involved. The entropy of the system as a first derivative of the Gibbs energy with respect to the temperature, undergoes a jump. First, let us to have a look at the general behavior of the Gibbs function against the horizon radius of the black hole.
In canonical ensemble, the Gibbs free energy is obtained through the relation $G=H-TS$, which in our case leads to
\begin{equation}\label{gr}
\begin{split}
G&=\frac{\left(-8 r_{h}^{4} \pi  P -\left(a +1\right) r_{h}^{2}+Q^{2}+g \right) \left(4 \ln \! \left(r_{h}\right) g +r_{h}^{2}\right)}{4 r_{h} \left(r_{h}^{2}+2 g \right)}+\\
&\frac{4 r_{h}^{3} \pi  P}{3}+\frac{r_{h} a}{2}+\frac{g}{2 r_{h}}+\frac{r_{h}}{2}+\frac{Q^{2}}{2 r_{h}}.
\end{split}
\end{equation} 
\begin{figure}[h!]
\centering
\subfloat[a]{\includegraphics[width=5cm]{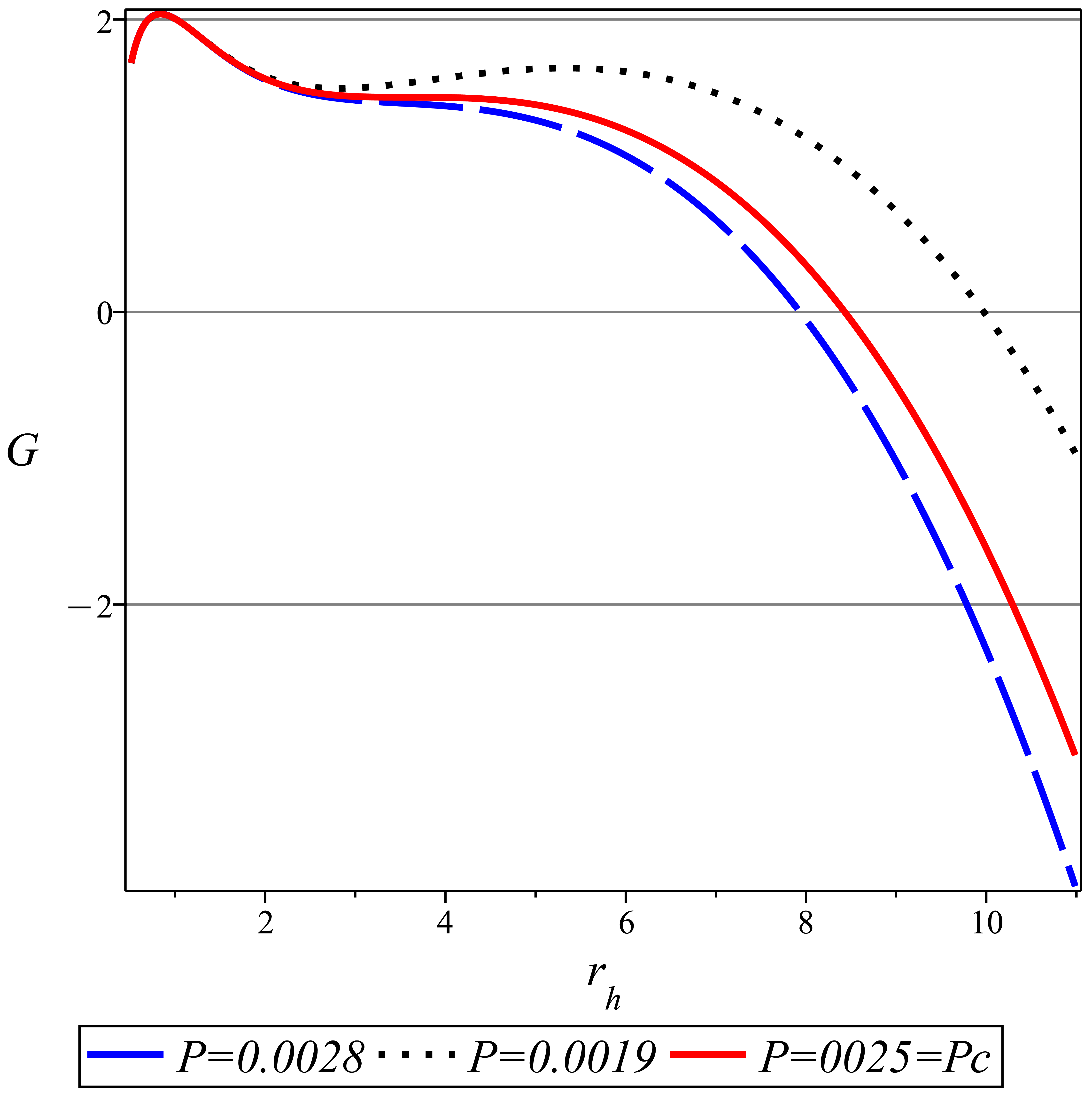}}\qquad
\subfloat[b]{\includegraphics[width=5cm]{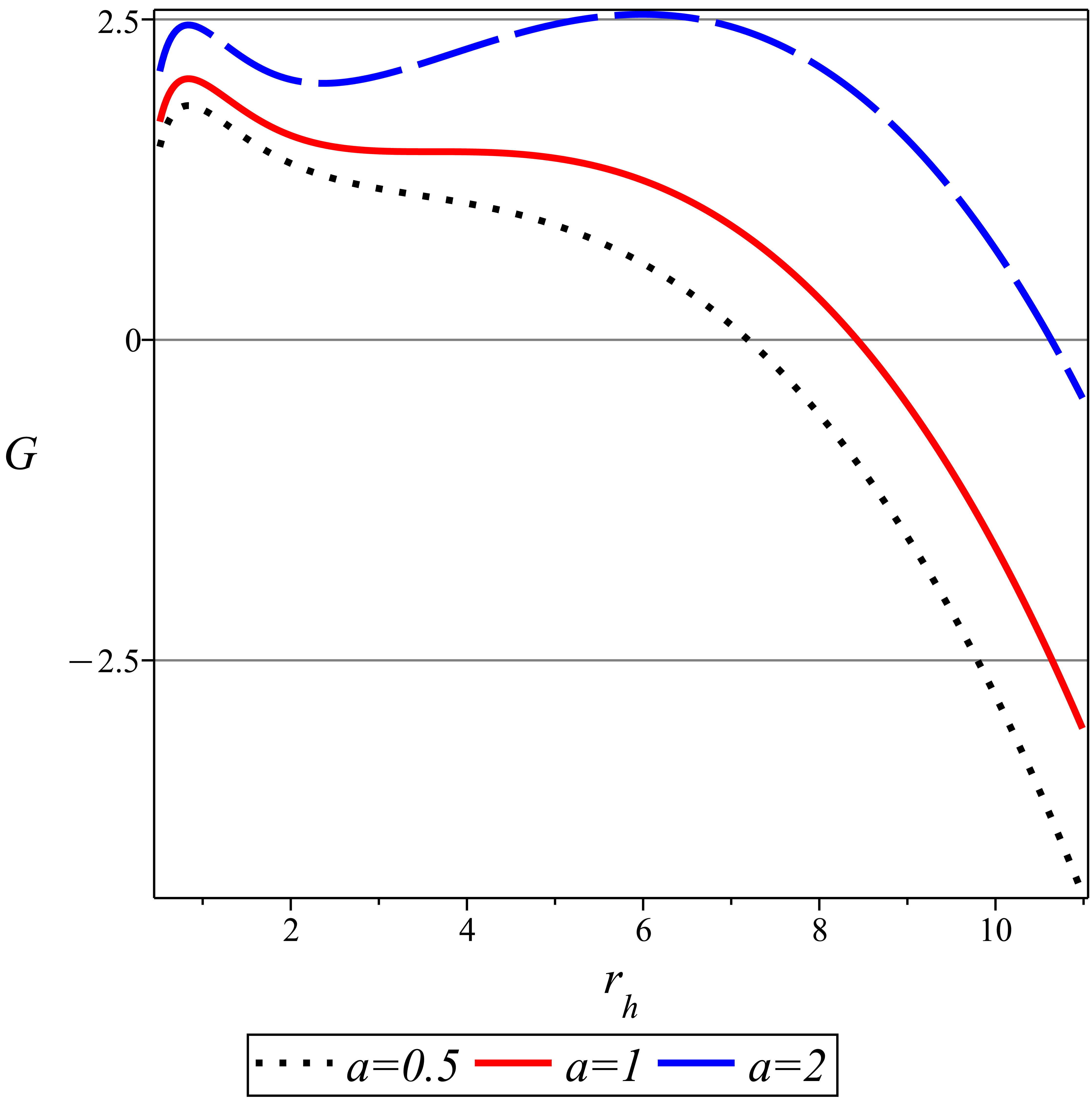}}
\caption{$G-r_{h}$ diagram of the system; \textbf{(a)}: for different values of the pressure and \textbf{(b)}: when the cloud of strings parameter changes and pressure is fixed at the value $P=0.0025$. \label{fig:5}}
\end{figure}
\begin{figure}[h!]
\centering
\subfloat[a]{\includegraphics[width=5cm]{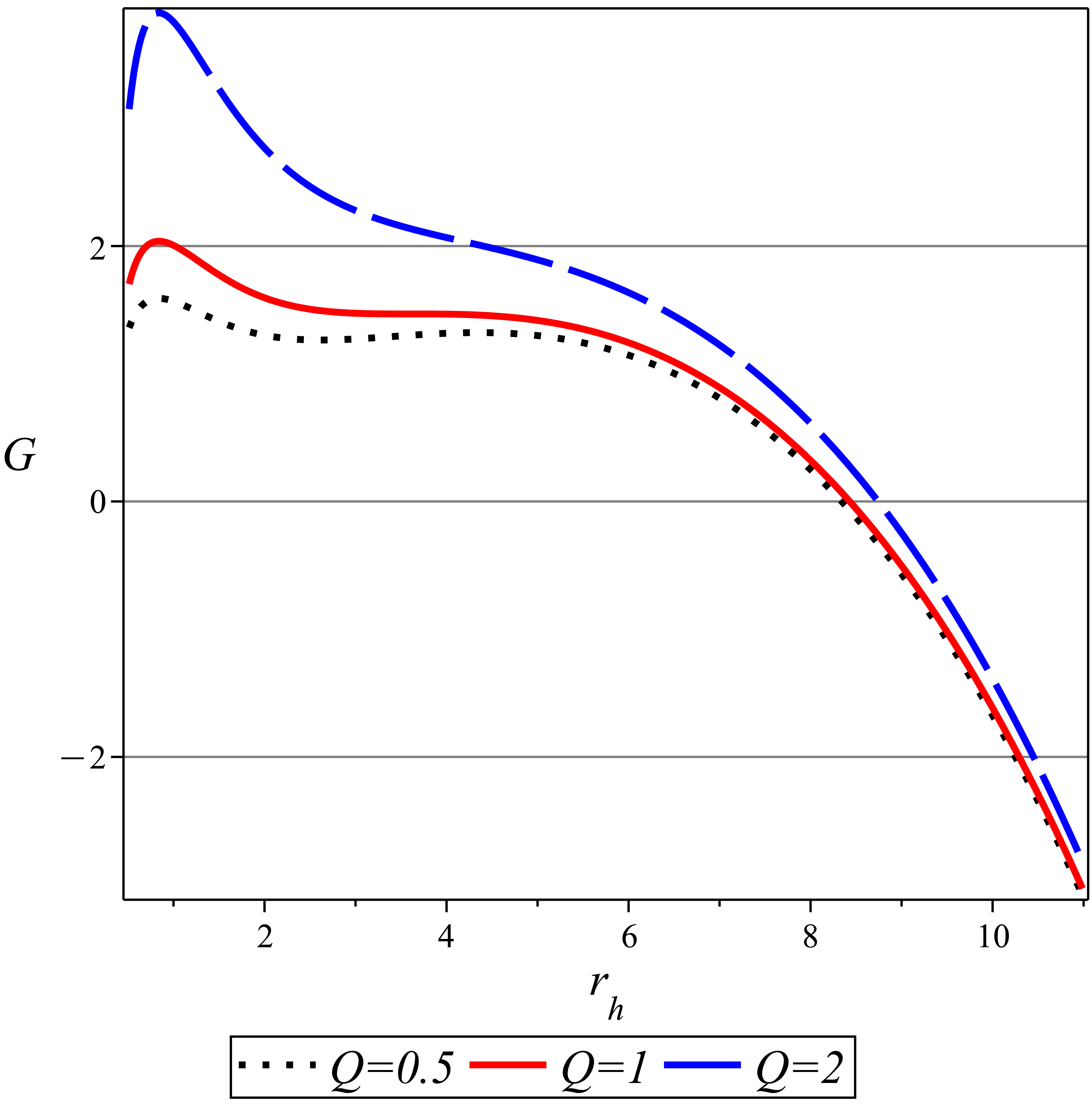}}\qquad
\subfloat[b]{\includegraphics[width=5cm]{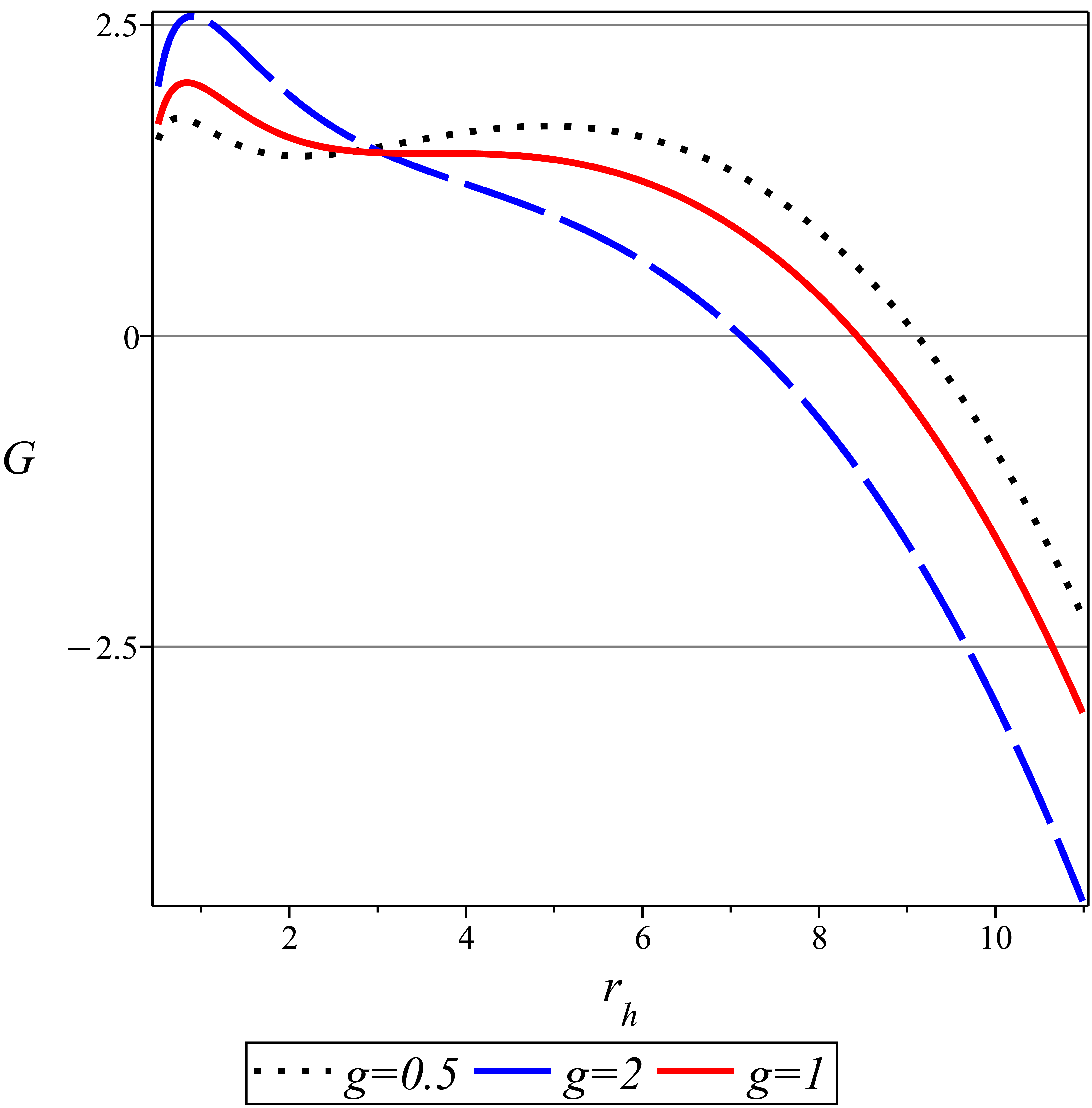}}
\caption{$G-r_{h}$ diagram of the system; \textbf{(a)}: when the Yang-Mills charge changes, \textbf{(b)}: when the GB parameter changes. \label{fig:6}}
\end{figure}
In Figs (\ref{fig:5}) and (\ref{fig:6}) the behavior of the Gibbs function for different values of parameters is seen. In all diagrams an inflection point is seen which indicates a phase transition. Unstable and stable regions are defined by conditions $G>0$ and $G<0$ respectively. The left panel of Fig. (\ref{fig:5}) shows that as the pressure increases, the Gibbs function cuts the horizon axis in smaller distances and vice-versa.  The local minimums are above the horizon axis, where $G>0$. The Global stable states occur for large horizons and bellow the horizon axis where $G<0$.  \par
\begin{figure}[h!]
\centering
\subfloat[a]{\includegraphics[width=5cm]{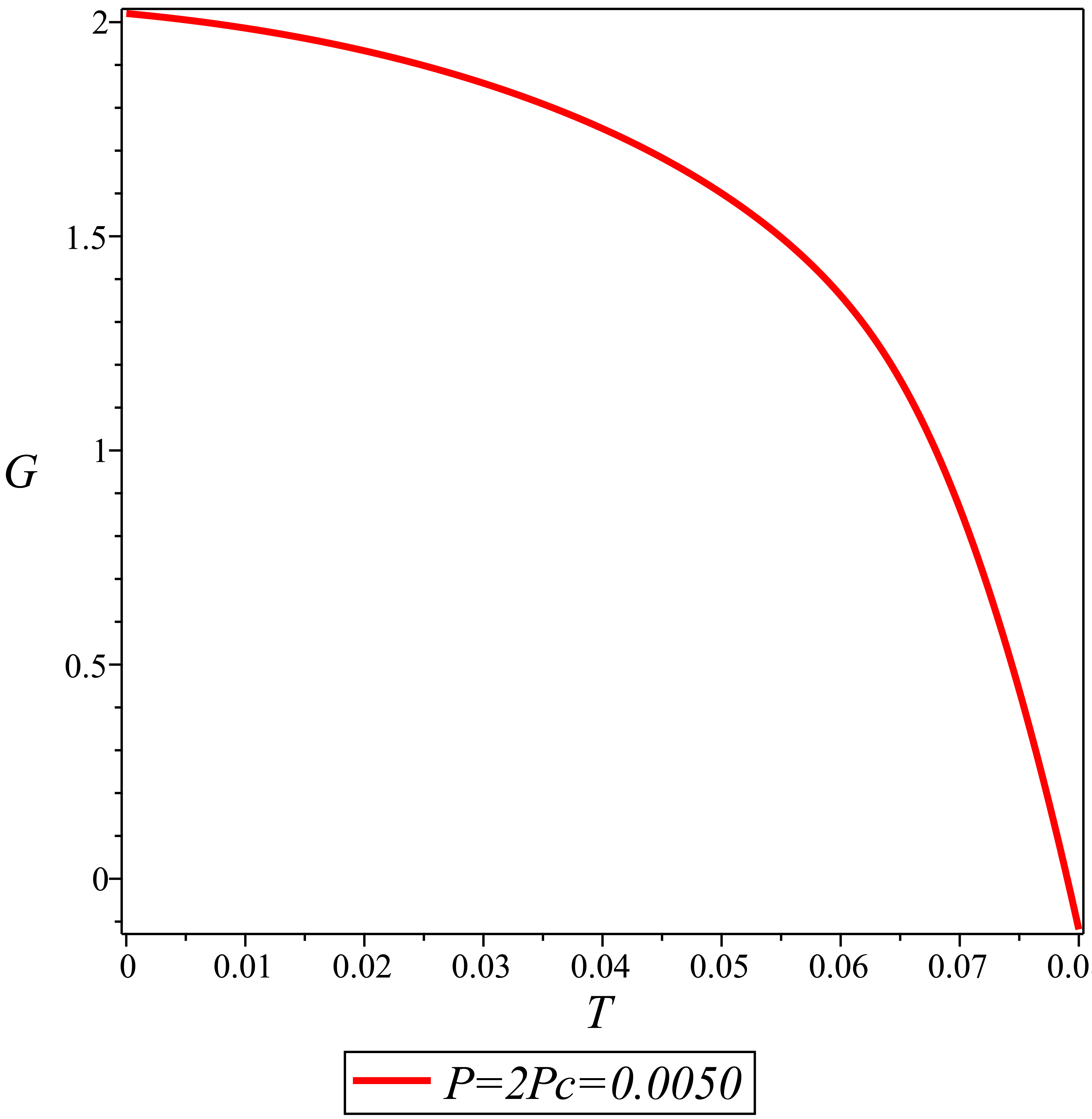}}\quad
\subfloat[b]{\includegraphics[width=5cm]{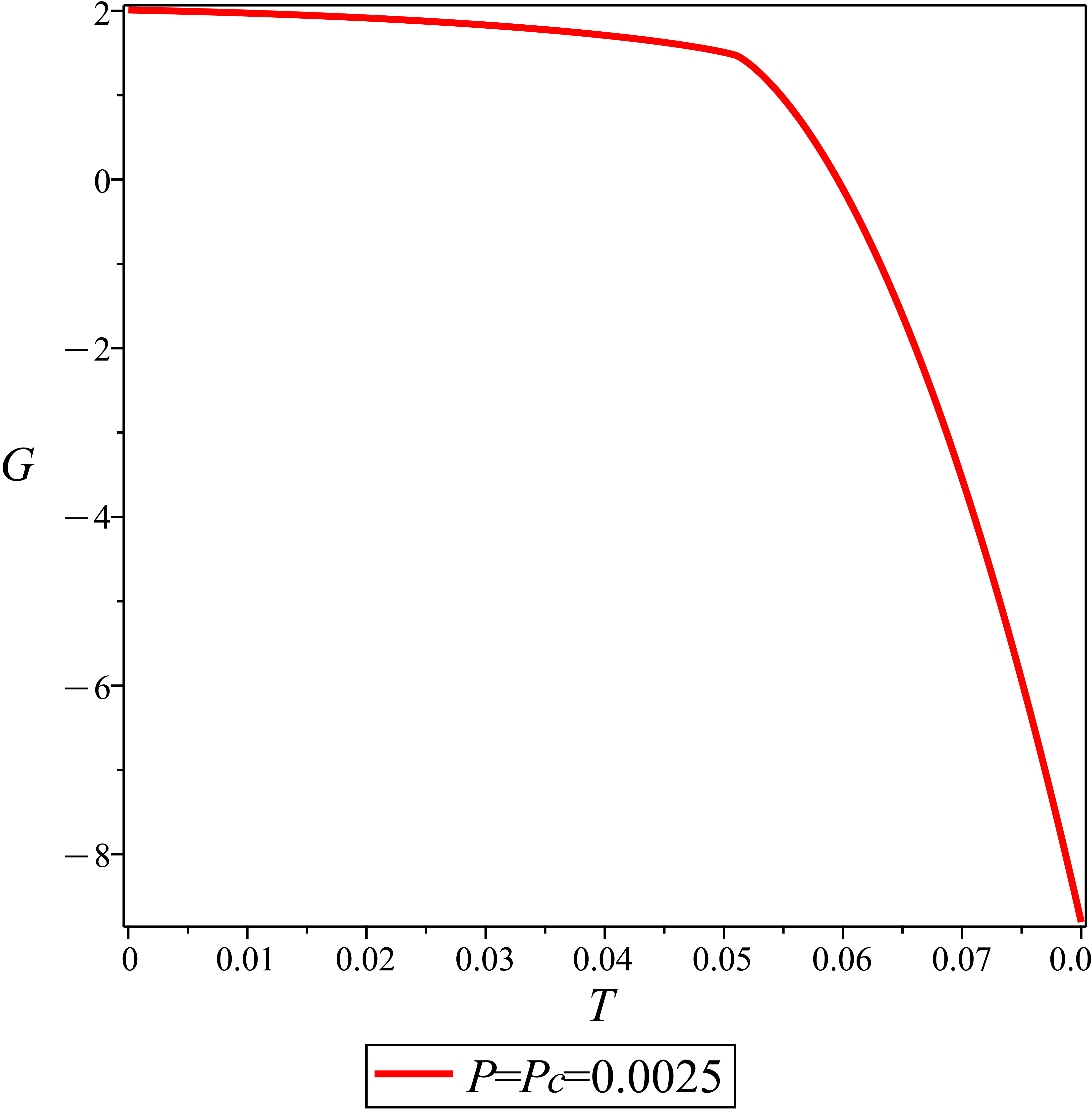}}\quad
\subfloat[b]{\includegraphics[width=5cm]{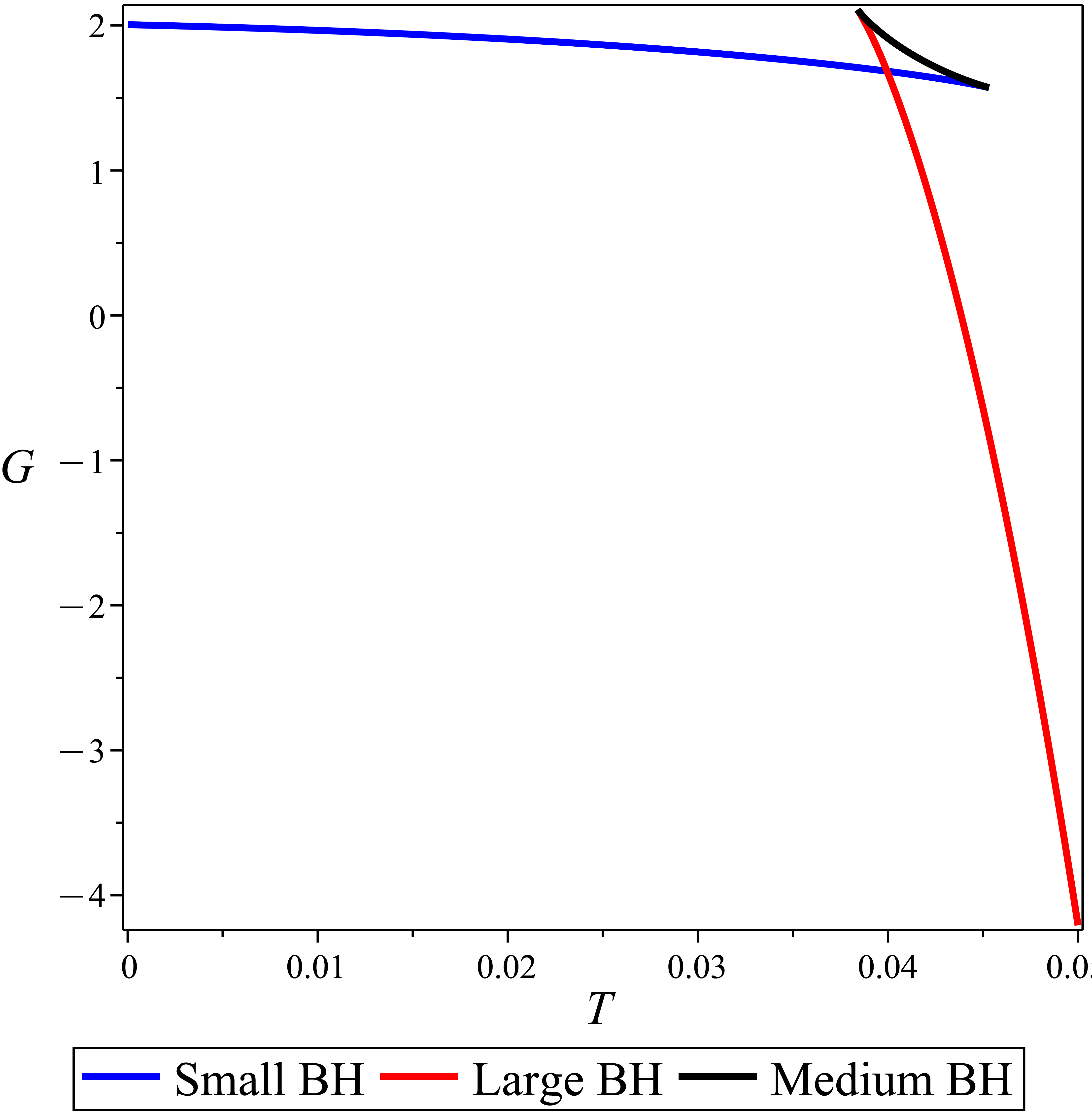}}
\caption{$G-T$ diagram of the system; \textbf{(a)}: for the pressure higher than the critical pressure,\textbf{(b)}: for the critical pressure and \textbf{(c)}: for the pressure $P=\frac{P_c}{2}$ less than the critical pressure. The critical pressure $P_c=0.0025$ corresponds to the values of system parameters $a=1,g=1, Q=1$. A swallow-tail shape is seen in the right panel. \label{fig:7}}
\end{figure}

 
Now, by using the equation of state (\ref{state}) and equation (\ref{gr}), the behavior of the Gibbs function versus temperature is investigated. In Fig. (\ref{fig:7}), the left panel shows the Gibbs function for the pressure higher than the critical pressure and a smooth behavior is seen. The middle panel represents the Gibbs function for the critical pressure for the values $a=1, g=1$ and $Q=1$ where a kink is seen in the diagram. The right panel, represents the swallow-tail behavior of the system for the pressure less than the critical pressure. In these diagrams, the critical pressure is $P_c=0.0025$ which is associated to the values $a=1,g=1$ and $Q=1$.
\begin{figure}[h!]
\centering
\subfloat[a]{\includegraphics[width=5cm]{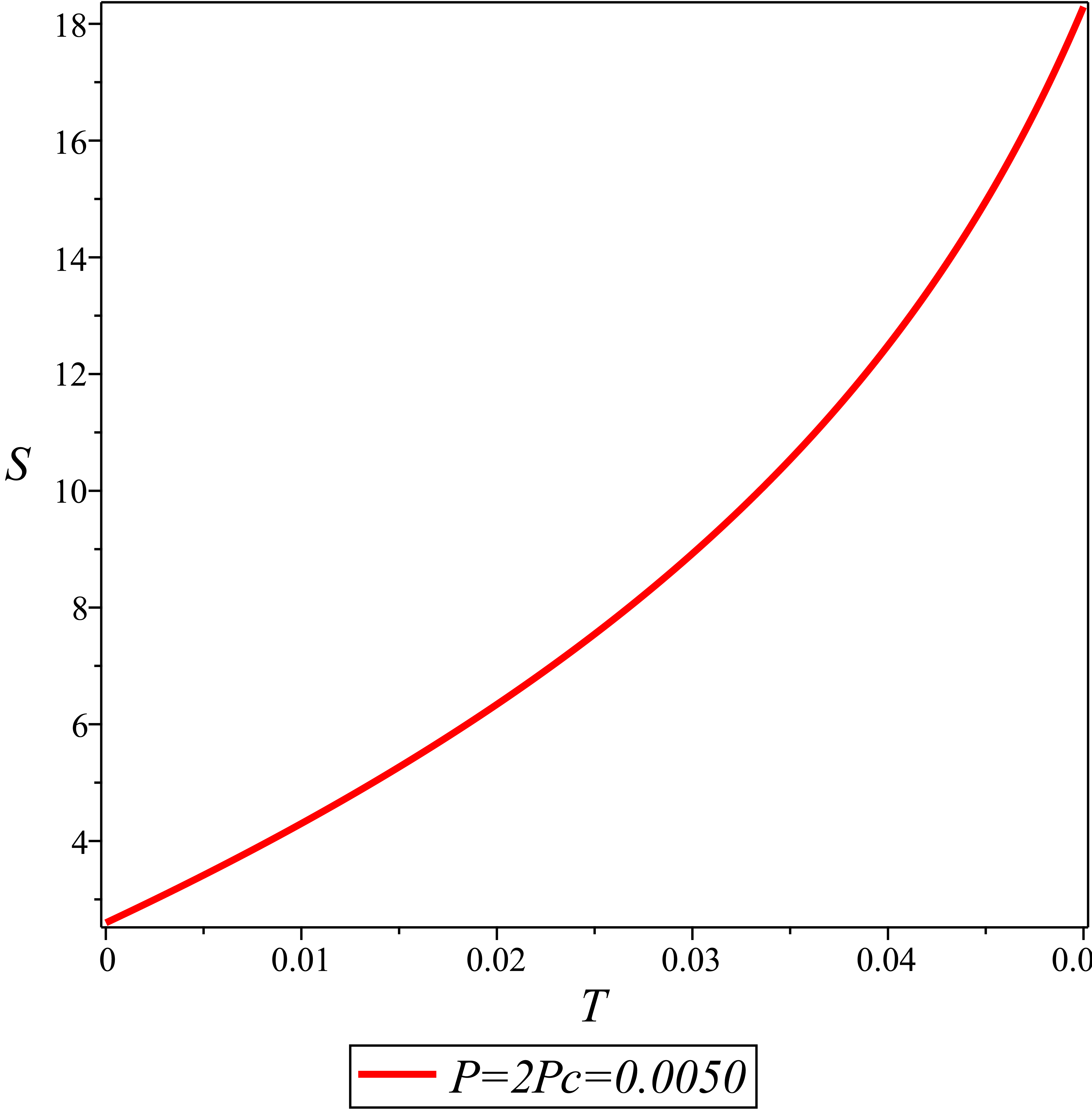}}\quad
\subfloat[b]{\includegraphics[width=5cm]{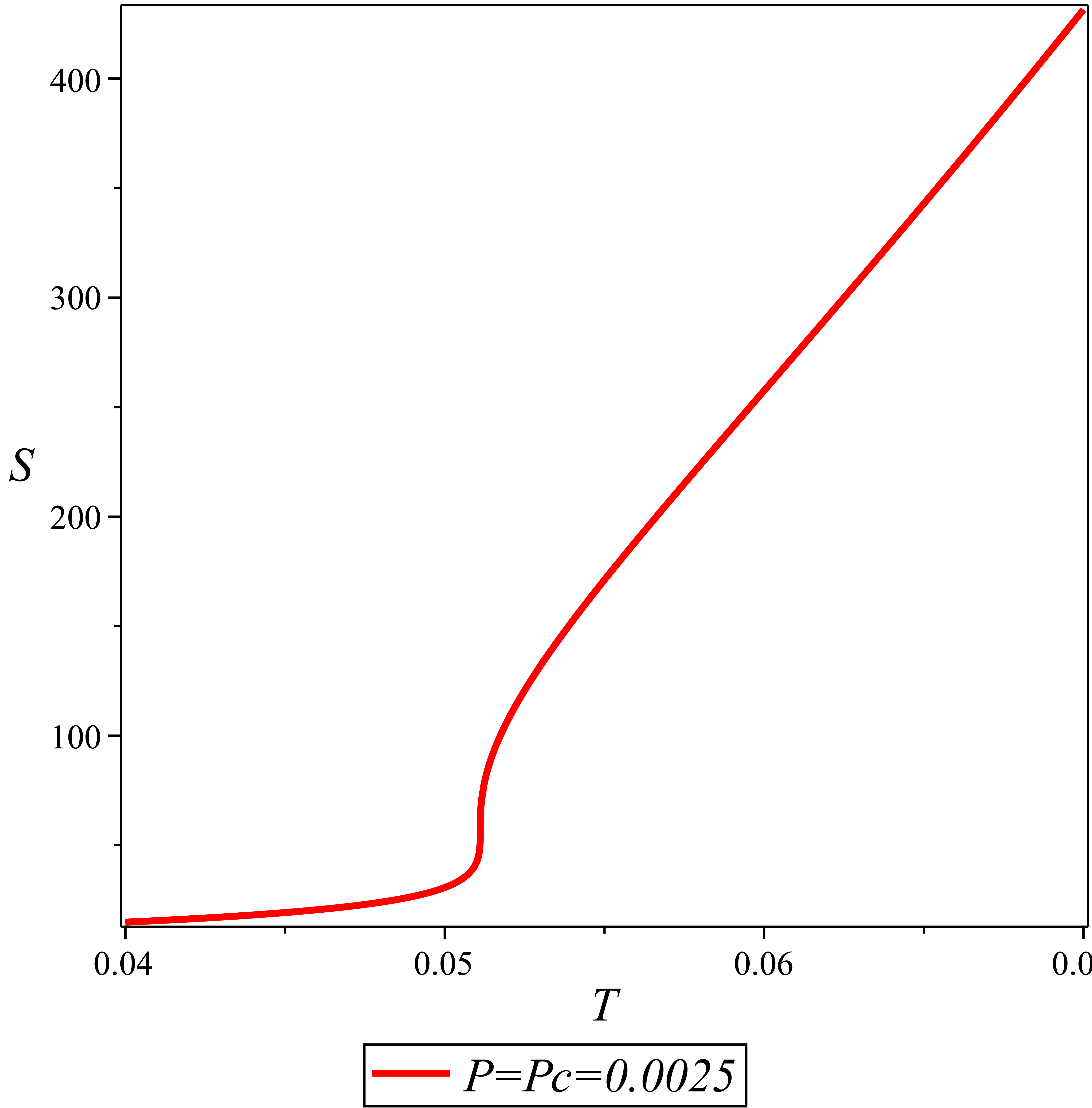}}\quad
\subfloat[b]{\includegraphics[width=5cm]{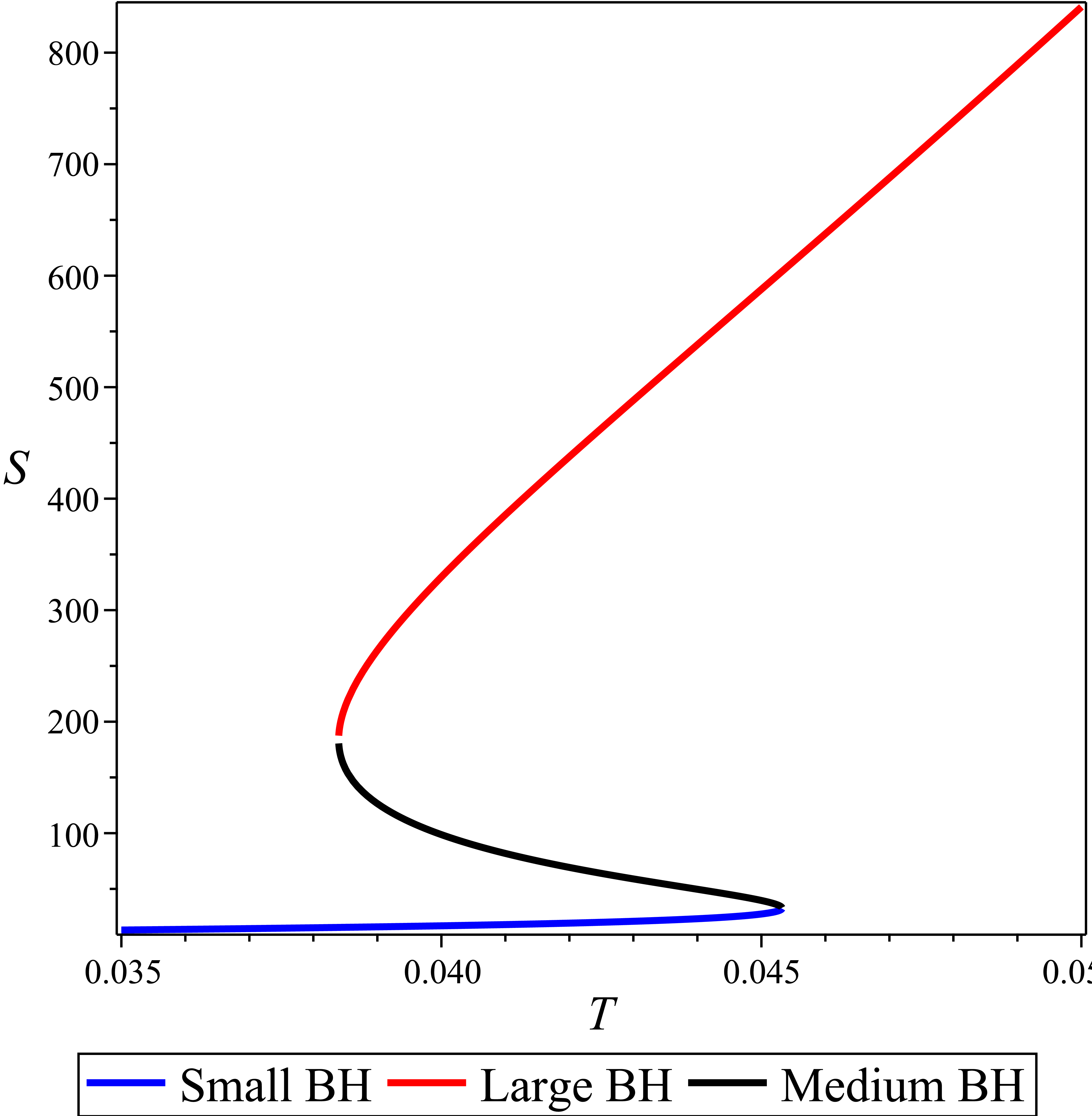}}
\caption{$S-T$ diagram of the system; \textbf{(a)}: for the pressure higher than the critical pressure,\textbf{(b)}: for the critical pressure and \textbf{(c)}: for the pressure $P=\frac{P_c}{2}$ less than the critical pressure . \label{fig:8}}
\end{figure}
The behavior of the entropy versus temperature can be seen in Fig. (\ref{fig:8}). The left panel shows the entropy at the pressure higher than the critical pressure. The middle and right panels show the entropy for pressures equal to critical pressure and lower than the critical pressure respectively. The behavior of these diagrams also indicates to a phase transition.\par 
Now, we study the critical exponents of the system to compare with the van-der walls fluid to determine the behavior of physical quantities near the critical point. The relation between thermodynamic quantities and critical exponents $\alpha, \beta, \gamma$ and $\delta$ are as follow,
\begin{equation}
C_v= \left(T \frac{\partial S}{\partial T}\right)_{v} \propto \vert t \vert ^{-\alpha},
\end{equation}
\begin{equation}
\eta=v_l-v_s \propto \vert t \vert ^{\beta},
\end{equation}
\begin{equation}
\kappa_T=-\frac{1}{v} \left(\frac{\partial v}{\partial p} \right)_{T} \propto \vert t \vert ^{-\gamma}
\end{equation}
and
\begin{equation}
\vert P-P_c \vert \propto \vert v-v_c \vert^{-\delta}.
\end{equation}
Where, $C_v,\eta$ and $\kappa_T$ are heat capacity at constant volume, the order parameter and the isothermal compressibility
respectively. The expression $\vert P-P_c \vert$ describes the behavior of critical isotherms.
Usually, the equation of state is transformed to an equation which is called law of corresponding state. In thermodynamics, the equation of state is written in this way to investigate the behavior of gases near the critical point and independent of the type of gas. In this regard, we change the form of equation of state by using rescaled quantities $\nu=\frac{v}{v_c}, \tau=\frac{T}{T_c}$ and $p=\frac{P}{P_c}$ where $v=2r_h$ is specific volume in four dimensions. Now, the equation of state transforms to the law of corresponding state
\begin{equation}\label{lc}
\displaystyle p =\frac{T_{c}}{ v_{c} p_{c}} \frac{\tau}{\nu}+\frac{8 g T_{c}}{ v_{c}^{3} p_{c}}\frac{\tau}{\nu^{3}}-\frac{a +1}{2 \pi  v_{c}^{2} p_{c}} \frac{1}{\nu^{2}}+\frac{2 Q^{2}+2g}{\pi  v_{c}^{4} p_{c}} \frac{1}{\nu^{4}}.
\end{equation}
In order to study the quantities near the critical points, we substitute relations $\tau=1+t$ and $\nu=1+\omega$ into equation (\ref{lc}). The law of corresponding state is approximated as
\begin{equation}\label{ex}
p=1+\Theta t+ \Phi t \omega+\Omega \omega^3+\cdots
\end{equation}
where, 
\begin{equation}
\Theta=\frac{T_{c} \left(v_{c}^{2}+8 g \right)}{v_{c}^{3} p_{c}},
\end{equation}
\begin{equation}
\Phi=-\frac{T_{c} \left(v_{c}^{2}+24 g \right)}{v_{c}^{3} p_{c}}
\end{equation}
and
\begin{equation}
\Omega= -\frac{\pi  T_{c} v_{c}^{3}+80 \pi  g T_{c} v_{c}-2 a v_{c}^{2}+40 Q^{2}-2 v_{c}^{2}+40 g}{\pi  v_{c}^{4} p_{c}}.
\end{equation}
The terms proportional to $\omega$ and $\omega^2$ are not present in this expansion. Because, their associated coefficients $\frac{\partial p}{\partial \omega}$ and $\frac{\partial^2 p}{\partial \omega^2}$ at critical point must be equal to zero.
Since, the entropy $S$, does not depend on temperature $T$, so, $C_v$ is equal to zero and this gives $\alpha=0$. To estimate the critical exponent $\beta$, we must evaluate the specific volumes of large and small black holes, i.e. $v_l$ and $v_s$. During the phase transition the pressure of the black hole remains unchanged. So, from relation (\ref{ex}), we get
\begin{equation}\label{eq}
1+\Theta t+ \Phi t \omega_l+\Omega \omega^3_l=1+\Theta t+ \Phi t \omega_s+\Omega \omega^3_s.
\end{equation}
On the other hand, during the phase transition, Maxwell's equal area law, says
\begin{equation}
\int_{\omega_s}^{\omega_l} \omega \frac{dp}{d\omega}d\omega=0
\end{equation}
where, in this case leads to
\begin{equation}
\Phi t(\omega_l^2-\omega_s^2)=-\frac{3}{2}\Omega (\omega_l^4-\omega_s^4).
\end{equation}
Now, by using this equation and  relation (\ref{eq}), one can get
\begin{equation}
\omega_l=-\omega_s=\sqrt{-\frac{\Phi t}{\Omega}}.
\end{equation}
Thus, the order parameter can be obtained as
\begin{equation}
\eta=v_l-v_s=v_c (\omega_l-\omega_s)=2v_c\omega_l \propto \sqrt{-t}.
\end{equation}
This leads to the result $\beta=\frac{1}{2}$.
The isothermal compressibility can be evaluated as follows
\begin{equation}
\kappa_T=-\frac{1}{v_c(1+\omega)}\frac{\partial v}{\partial \omega}\left(\frac{\partial \omega}{\partial P}\right)_T \propto \left(-\frac{1}{\frac{\partial p}{\partial \omega}}\right)_{\omega=0}=-\frac{t^{-1}}{\Phi}
\end{equation}
Which gives $\gamma=1$.
The critical isotherm can be investigated by using relation (\ref{ex}) at $t=0$( $T=T_c$). This gives
\begin{equation}
p=1+\Omega \omega^3
\end{equation}
which leads to the result $\delta=0$. These critical exponents which are close to the critical exponents of the van-der Walls fluid, satisfy the following relations:
\begin{equation}
\begin{split}
\alpha+2 \beta+\gamma &=2,\quad \alpha+\beta(\delta+1)=2,\quad (2-\alpha)=\delta(1-\alpha)-1\\
\gamma &=\beta(\delta-1),\quad \gamma(\delta+1)=(\delta-1)(2-\alpha)
\end{split}
\end{equation}
According to this investigation we can conclude that, the thermodynamics exponents associated with $4D$ AdS-GB-YM black hole in the presence of a cloud of strings satisfies the mean field theory prediction.\par
\begin{figure}[h!]
\centering
\subfloat[a]{\includegraphics[width=5cm]{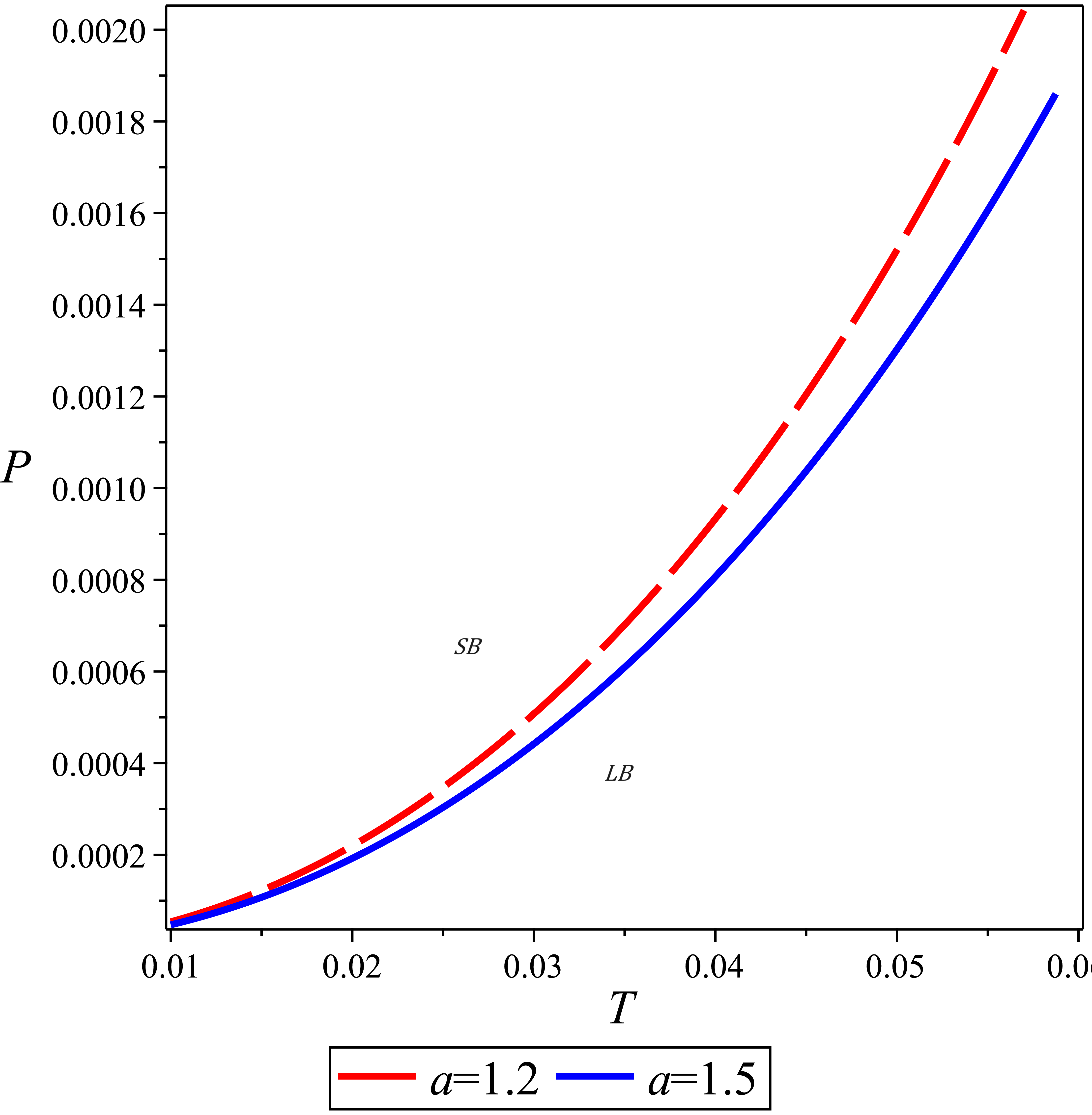}}\quad
\subfloat[b]{\includegraphics[width=5cm]{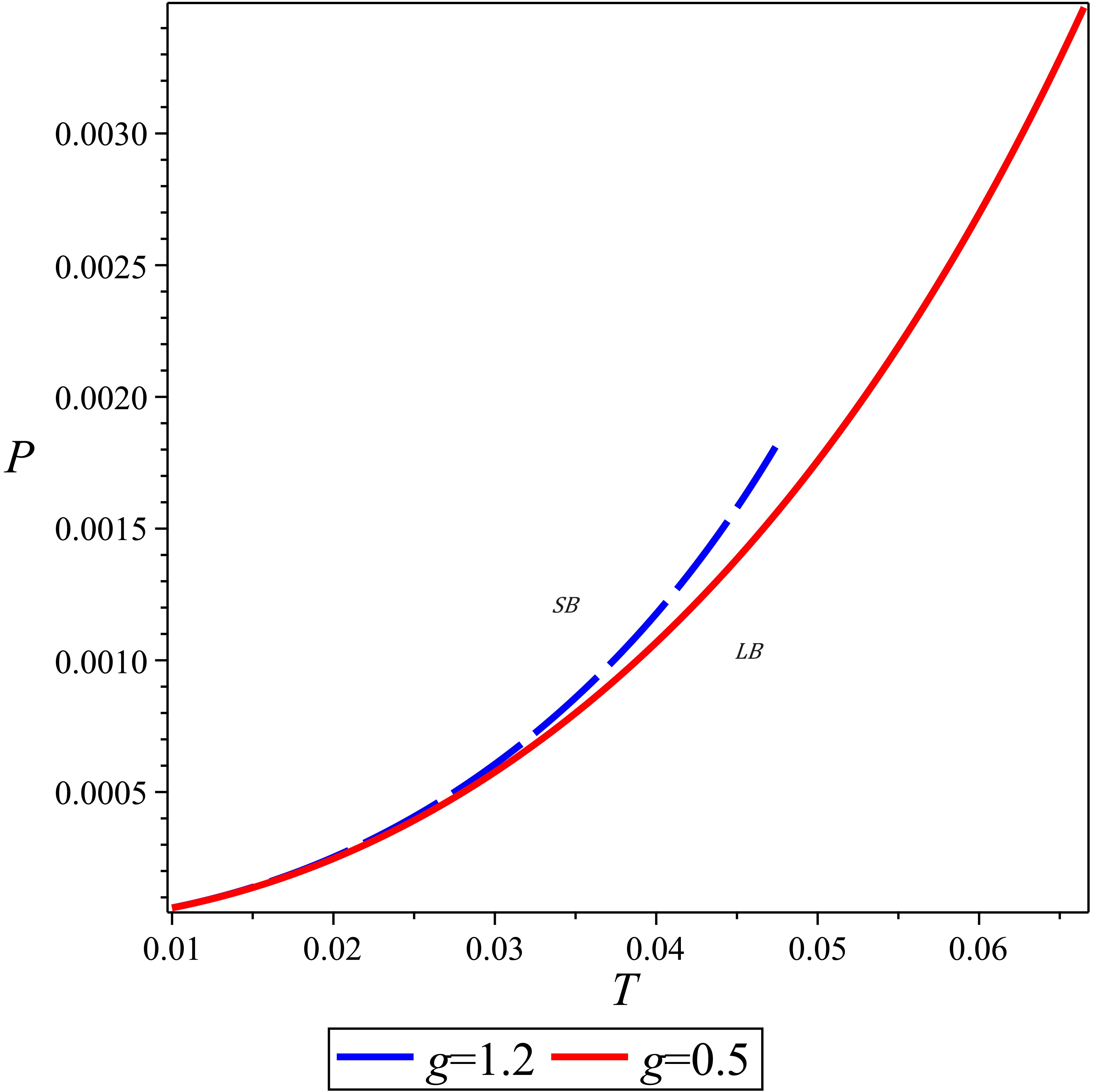}}\quad
\subfloat[b]{\includegraphics[width=5cm]{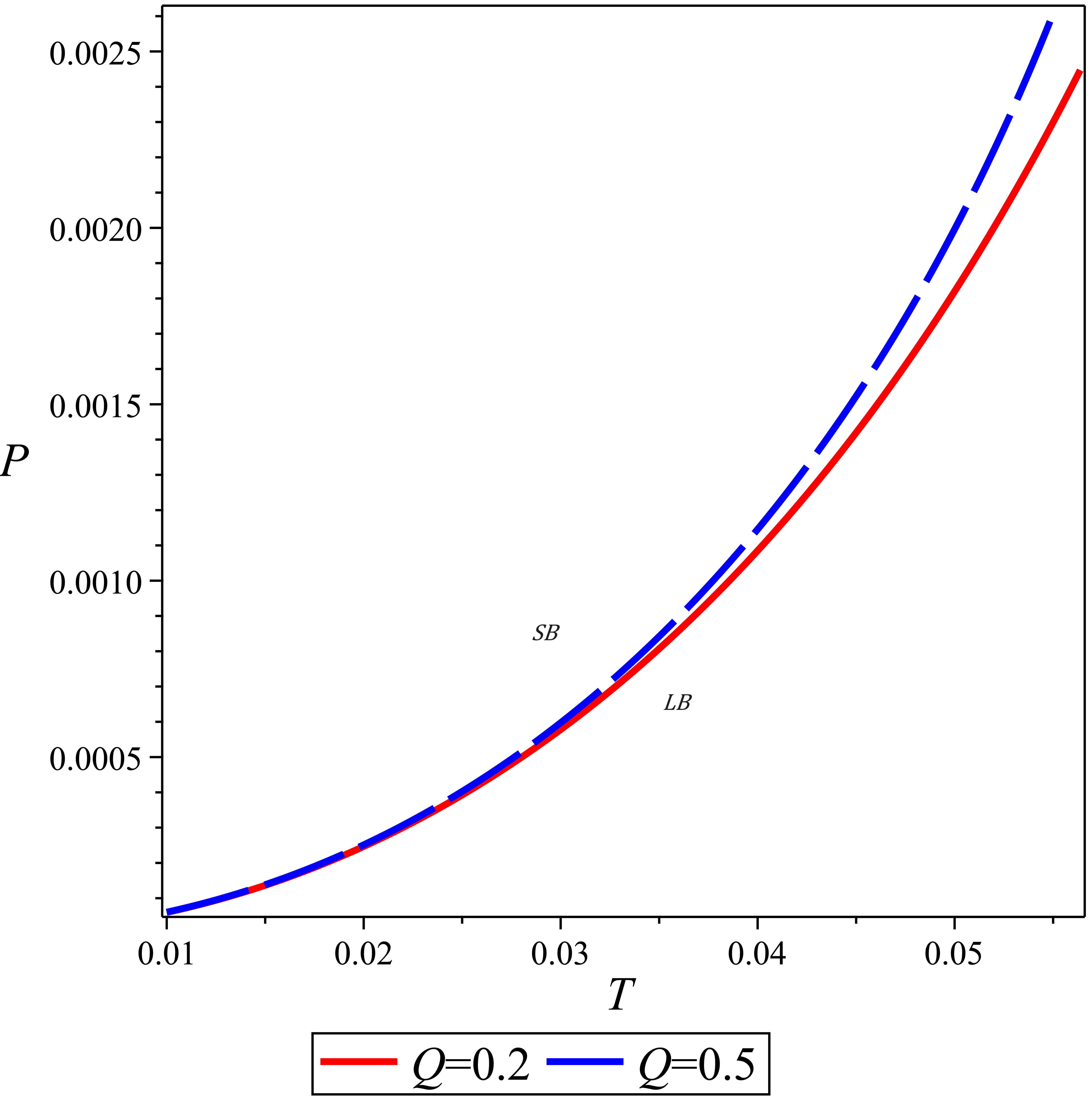}}
\caption{The co-existence curve of the system; \textbf{(a)}: for different values of the cloud of strings parameter, \textbf{(b)}: for different values of the GB parameter and \textbf{(c)}: for different values of the Yang-Mills charge. These curves are the boundary between the small and large black hole phases. \label{fig:9}}
\end{figure}

 
In this step, we investigate the coexistence curves of this model. The notion of co-existence curve plays an essential role in the study of the first order phase transition of the system which is usually accompanied by a latent heat. The co-existence curve is the boundary between two different phases in $P-T$ plane on which the Gibbs function values of both phases are the same. In our system one of the phases is associated to small black holes and another one is the large black holes. The slope of the co-existence curve is obtained through this fact that on the co-existence curve in the $P-T$ plane the value of the Gibbs function remains unchanged at any point. This leads to the famous Clapeyron equation which gives the slope of the co-existence curve in the $P-T$ plane. The Clapeyron equation can be written as follows
\begin{equation}
\frac{dP}{dT}= \frac{T\Delta S}{\Delta V}=\frac{(\Delta H)_P}{\Delta V}
\end{equation}
Where, $\Delta S$ and $\Delta V$ denote the entropy and volume change between two phases. Also, $(\Delta H)_P$ is the enthalpy change during the phase transition at constant pressure. Here, we will not use the Clapeyron equation directly in our analyses. Rather, we use equations (\ref{state}) and (\ref{gr}) for fixed values of the Gibbs function and by removing $r_h$ from these equations, we obtain the co-existence curves in the $P-T$ plane.
Fig. (\ref{fig:9}) shows the co-existence curve for some arbitrary values of parameters. The behavior of curves is in agreement with the behavior of co-existence curves of a van-der walls fluid during the first order phase transition. 
Small black holes live above the co-existence curve, and large black holes live below it.\par
\begin{figure}[h!]
\centering
\subfloat[a]{\includegraphics[width=5cm]{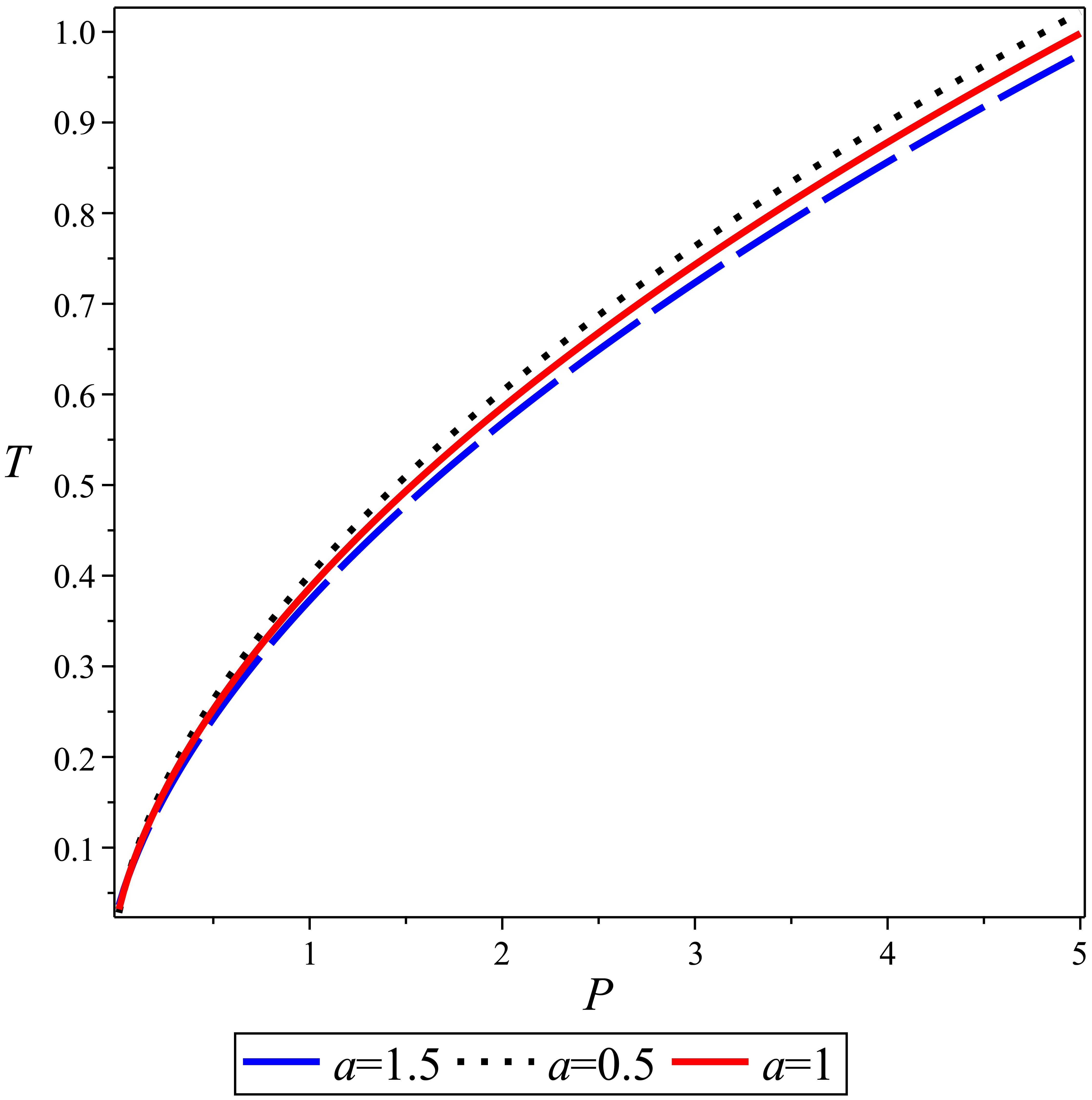}}\quad
\subfloat[b]{\includegraphics[width=5cm]{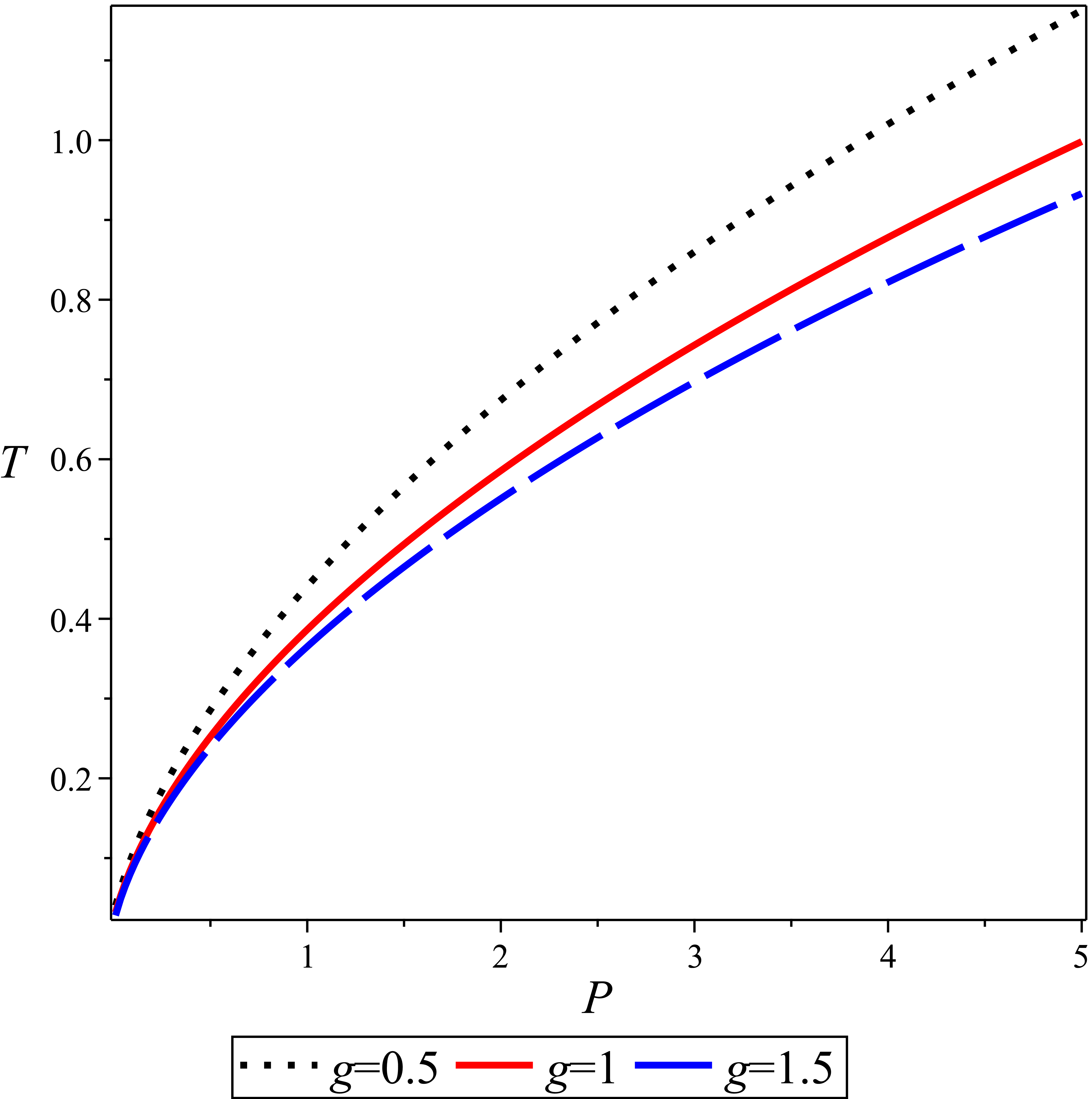}}\quad
\subfloat[b]{\includegraphics[width=5cm]{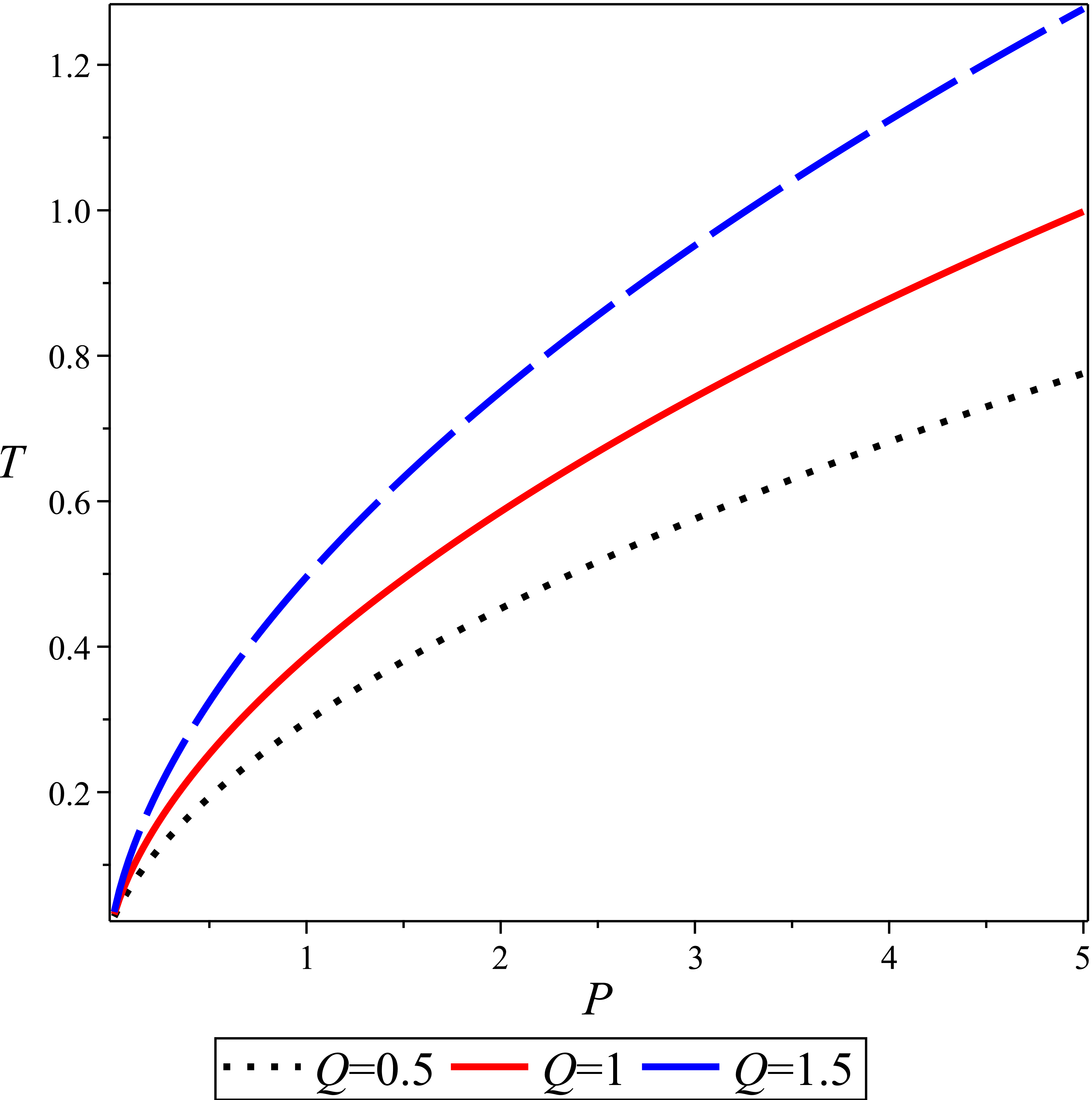}}
\caption{The inversion curves for different values of involved parameters; \textbf{(a)}: the cloud of strings parameter changes, \textbf{(b)}: the GB parameter changes and \textbf{(c)}: the Yang-Mills charge changes. \label{fig:10}}
\end{figure}

Finally, we investigate the Joule-Thomson expansion in this model. During the Joule-Thomson expansion a thermodynamic system experiences heating or cooling at constant enthalpy \cite{oz}. The criterion for checking the cooling or heating of a black hole is the Joule-Thomson coefficient
\begin{equation}
\mu =\left(\frac{\partial T}{\partial P}\right)_{H}.
\end{equation}
The sign of the $\mu$ determines whether cooling ($\mu>0$) or heating ($\mu <0$) will occur.
For $\mu=0$, the associated temperature is called inversion temperature $T_i$ and its associated curve determines the boundary between the heating and cooling regions in $T-P$ plane.
The general relation of Joule-Thomson coefficient is as bellow \cite{oz}
\begin{equation}\label{mu}
\mu=\frac{1}{C_P}\left[ T \left( \frac{\partial V}{\partial T}\right)_P -V \right].
\end{equation}
The condition $\mu=0$ gives 
\begin{equation}\label{inv}
T_i=V \left( \frac{\partial T}{\partial V}\right)_P.
\end{equation}
By using this equation and Hawking temperature in relation (\ref{ht}), one can depict the inversion curves. In Fig.(\ref{fig:10}), the behavior of inversion curves for several values of parameters is seen.
\begin{figure}[h!]
\centering
\subfloat[a]{\includegraphics[width=5cm]{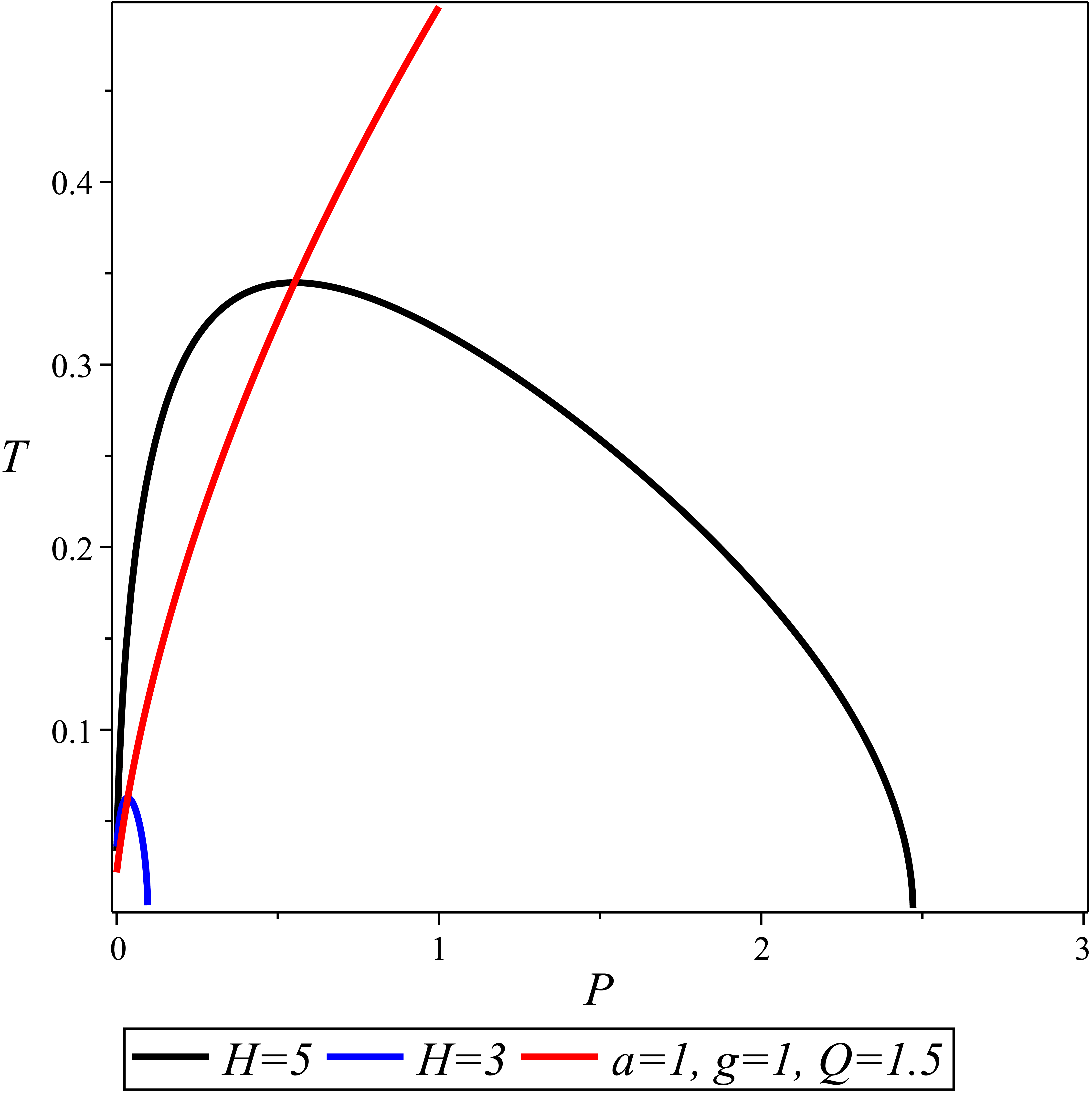}}\quad
\subfloat[b]{\includegraphics[width=5cm]{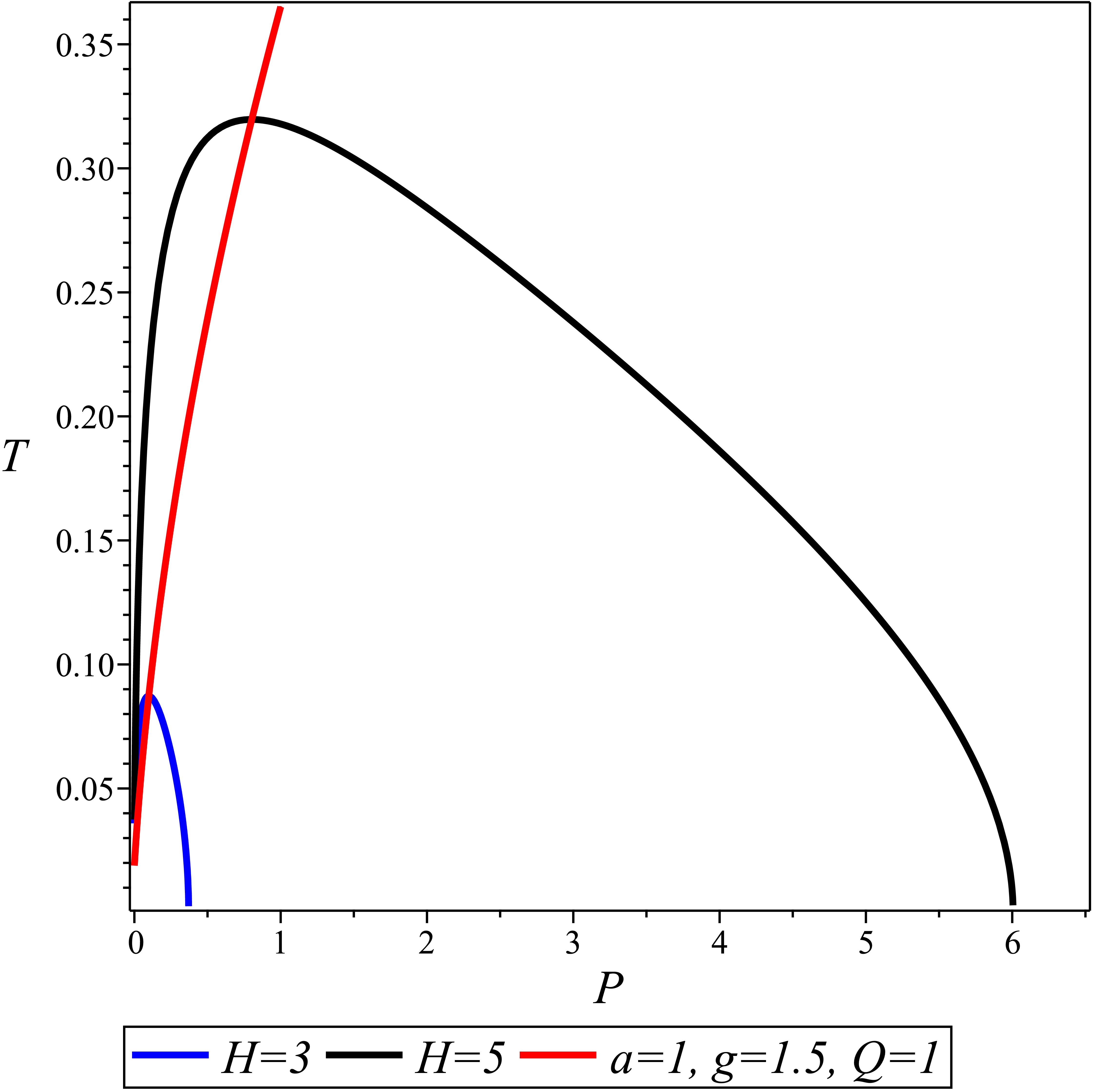}}\quad
\subfloat[b]{\includegraphics[width=5cm]{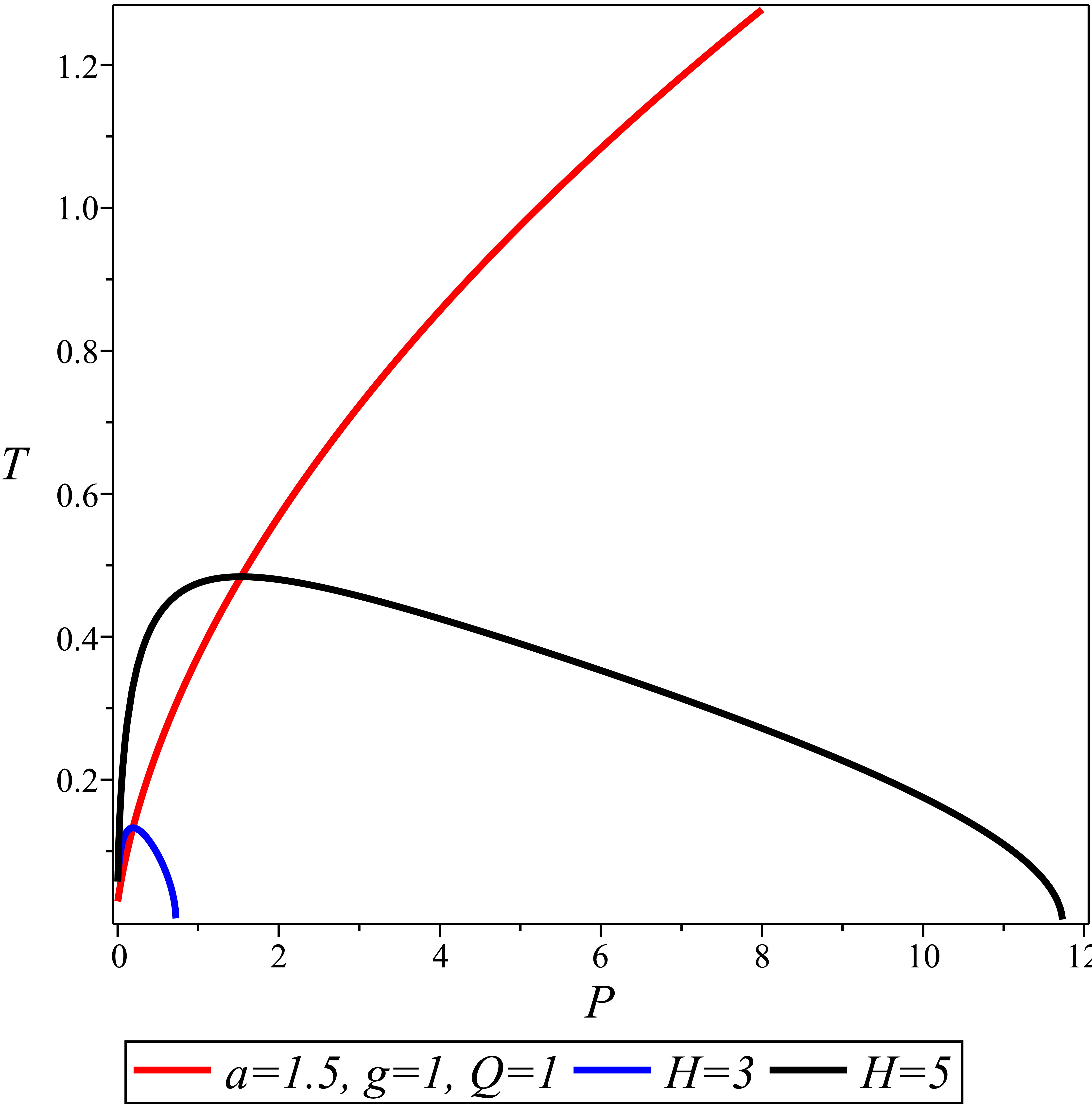}}
\caption{Inversion and isenthalpic curves for different values of parameters. In all panels, the blue and black curves are the isenthalpic curves with the values $H=3$ and $H=5$ respectively. The inversion curves have been depicted for \textbf{(a)}: $a=1, g=1, Q=1.5$, \textbf{(b)}: $a=1, g=1.5, Q=1$ and \textbf{(c)}: $a=1.5, g=1, Q=1$.   \label{fig:11}}
\end{figure}
Now, by using equations (\ref{e1}) and (\ref{ht}), we depict the isenthalpic curves in the $T-P$ plane. See Fig. (\ref{fig:11}). The inversion curve in each plot divides the $T-P$ plane to cooling and heating regions. Above the inversion curve, the Joule-Thomson coefficient $\mu$ is positive (cooling) and bellow it is negative(heating). The inversion curve intersects the isenthalpic curve where $\mu=0$. For the blue isenthalpic curves, $H=3$ while for the black curves $H=5$. These investigations  make us more confident about the existence of phase transition of the first type in this model. In order to study the thermodynamic behavior of black hole solution of various gravitational models, Refs. \cite{upa1}-\cite{upa15}, are suggested.
\newpage
\section{Conclusions}
The solution of the Yang-Mills charged AdS black hole in the context of $4D$ Einstein-Gauss-Bonnet(EGB) gravity in the presence of a cloud of strings was obtained. The necessary thermodynamic quantities were extracted to investigate the critical behavior and phase transition of the system. By examining the pressure of the system versus event horizon of the black hole, we realized that the system has critical behavior and its pressure diagrams are similar to the pressure diagrams of a Van der Waals fluid. The relations for critical values of the pressure, temperature and the volume of the system were obtained exactly in terms of the system parameters. During the investigation, we found that the effect of the cloud of strings parameter on some of the thermodynamic quantities is opposite to the effects of the Yang-Mills charge and the GB parameters. Necessary investigations showed that close to the critical temperature, the heat capacity diagram diverges, which can be a sign of the existence of the second type phase transition. The critical exponents for this system were obtained which satisfied the famous relations of the critical exponents of a Van der Walls fluid. Also, the examination of the Gibbs function showed that a swallow-tail shape is seen at the pressures lower than the critical pressure, which indicates a first order phase transition. Also, the investigation of the Clapeyron equation and its diagrams showed that there is a latent heat in this process and the phase transition between the small and large black hole can be considered as a first order transition. The Joule-Thomson expansion also was investigated and the heating and cooling regions of the system were identified.
According to these results and the dictionary of AdS/CFT correspondence, we can conclude that this system can be considered for the confinement-deconfinement phase transition and a phase transition from a normal phase to a superconducting phase in the dual conformal field on the boundary of the AdS space.

\vspace{1cm}
\noindent {\large {\bf Acknowledgment} }  Author would like to thank Dr. Özgür Ökcü for help in depicting the isenthalpic diagrams. 


\vspace{1cm}
\noindent {\large {\bf Data Availability statement } } \\\\
 All data that support the findings of this study are included within the article (and any supplementary
files).


\end{document}